\documentclass{aa}  
\usepackage{aalongtable}
\usepackage{graphicx}
\usepackage{txfonts}
\begin{document}
  \title{Dynamics of the NGC\,4636 Globular Cluster System
\thanks{ Based on observations collected at the European Southern Observatory, Cerro Paranal, Chile; ESO program  69.B-0366.}}
  \subtitle{An extremely dark matter dominated galaxy?}
    \author{Y.~Schuberth
    \inst{1,2}
    \and
    T.~Richtler 
    \inst{2} 
    \and B.~Dirsch 
    \inst{2} 
    \and M.~Hilker
    \inst{1} 
    \and S.~S.~Larsen 
    \inst{3} 
    \and M.~Kissler-Patig
    \inst{3}
    \and U.~Mebold 
    \inst{1}
  }
  
\offprints{{ylva@astro.uni-bonn.de}} \institute{Argelander-Institut
   f\"ur Astronomie \thanks{Founded by merging of the Institut f\"ur
   Astrophysik und Extraterrestrische Forschung, the Sternwarte, and
   the Radioastronomisches Institut der Universit\"at Bonn.},
   Universit\"at Bonn, Auf dem H\"ugel 71, D-53121 Bonn, Germany \and
   Universidad de Concepci\'on, Departamento de F\'isica, Casilla
   160-C, Concepci\'on, Chile \and European Southern Observatory,
   Karl-Schwarzschild-Str.~2, D-85748 Garching, Germany }
   \date{Received March --, 2005; accepted April 6, 2006}

 \abstract{
 We present the first dynamical study of the globular cluster system
   of NGC\,4636. It is the southernmost giant elliptical galaxy of the
   Virgo cluster, and is claimed to be extremely dark matter
   dominated, according to X--ray observations.}
{Globular clusters are used as dynamical tracers to investigate,
 by stellar dynamical means, the dark matter content of this
galaxy.}
   {Several hundred medium resolution spectra were acquired at the VLT
   with FORS\,2/MXU. We obtained velocities for 174 globular clusters
   in the radial range $0{\mbox{$.\!^{\prime}$}} 90 < R <
   15{\mbox{$.\!^{\prime}$}} 5 $, or $0.5\,- 9\,R_e$ in units of
   effective radius.  Assuming a distance of 15\,Mpc, the clusters are
   found at projected galactocentric distances in the range 4 to
   70\,kpc, the overwhelming majority within 30\,kpc. The measured
   line--of--sight velocity dispersions are compared to 
   Jeans--models.}
{ We find some indication for a
   rotation of the red (metal-rich) clusters about the minor axis.
   Out to a radius of 30\,kpc, we find a roughly constant projected
   velocity dispersion for the blue clusters of $\sigma \approx
   200\,\textrm{km\,s}^{-1}$.  The red clusters are found to have a
   distinctly different behavior: at a radius of about
   3{\mbox{$^\prime$}}, the velocity dispersion drops by
   $\sim\!50\,\textrm{km\,s}^{-1}\,$ to about
   $170\,\textrm{km\,s}^{-1}$\, which then remains constant out to a
   radius of 7{\mbox{$^\prime$}}.  The cause might be the steepening
   of the number density profile at $\sim\!3{\mbox{$^\prime$}}$
   observed for the red clusters.  Using only the blue clusters as
   dynamical tracers, we perform Jeans-analyses for different
   assumptions of the orbital anisotropy. Enforcing the model dark
   halos to be of the NFW type we determine their structural
   parameters. Depending on the anisotropy and the adopted M/L-values,
   we find that the dark matter fraction within one effective radius
   can vary between 20\% and 50\% with most a probable range between
   20\% and 30\%. A main source of uncertainty is the ambiguity of the
   velocity dispersion in the outermost bin.  A comparison with
   cosmological N--body simulations reveals no striking disagreement.}
   { Although the dark halo mass still cannot be strongly constrained,
   NGC 4636 does not seem to be extremely dark matter dominated.  The
   derived circular velocities are also consistent with Modified
   Newtonian Dynamics.}
  

\maketitle
%
\section{Introduction}
\subsection{Work on NGC\,4636}
NGC\,4636 is the southernmost bright elliptical in the Virgo cluster
of galaxies. It is located 10$^\circ$ south of M\,87, within the Virgo
Southern Extension. Surface brightness fluctuation (SBF) measurements
yield a distance of 15\,Mpc (Tonry et al.~\cite{2001ApJ...546..681T}),
which we adopt throughout this paper. This value, however, has to be
considered as a lower limit, since measurements based on the globular
cluster luminosity function (GCLF) presented by Kissler et
al.~(\cite{1994A&A...287..463K}) and Dirsch et al.~(\cite{boris}) point
towards a larger distance of more than 17\,Mpc. 
Table\,\ref{properties} summarizes the basic properties of NGC\,4636.

Although NGC\,4636, being an elliptical galaxy, predominantly consists
of an old stellar population, there are some weak indications of an
intermediate age population: firstly, it is a supernova Ia host galaxy
(SN 1939a: Zwicky \cite{1939PASP...51...36Z}, Giclas
\cite{1939PASP...51..166G}), and there are theoretical considerations
that this type of supernova originates in intermediate age populations
rather than in very old ones (Yungelson et
al.~\cite{1995ApJ...447..656Y} and Yoshii et
al.~\cite{1996ApJ...462..266Y}). Secondly, far infrared observations
by Temi et al.~(\cite{2003ApJ...585L.121T}) hint at a recent merger
which supplied NGC\,4636 with extra dust.

\begin{table}[b]
\caption[]{NGC\,4636 basic data compiled from the literature 
(1)\,NED,
(2)\,Tonry et al.~(\cite{2001ApJ...546..681T}), 
(3)\,this work,
(4)\,R3C, 
(5)\,Dirsch et al.~(\cite{boris}), 
(6)\,Bender et al.~(\cite{1994MNRAS.269..785B}),  
(7)\,Idiart et al.~({\cite{2003A&A...398..949I}}),
(8)\,Forman et al.~(\cite{1985ApJ...293..102F}),
(9)\,Loewenstein et al.~(\cite{2001ApJ...555L..21L})
.} 
\label{properties}

\begin{center}
\begin{tabular}{lll}
\hline
\hline
Other names &UGC\,07878, VCC\,1939&\\ \hline
Position (J2000) & $12\,\fh\, 42\, \fm \,50\, \fs$  $+02\, \fdg 41
\farcm 17 \farcs$& (1)\\
Galactic coordinates  &     $ l=297.75^\circ$  $b = 65.47^\circ$& \\
\hline
\hline
SBF distance modulus &$(m-M)= 30.83\pm 0.13$&(2)\\ 
Distance & $D=15\,\mathrm{Mpc}$&\\
Scale & 1\arcsec = 73\,pc$\quad$  1\arcmin = 4.4\,kpc&\\
GCLF distance modulus &$(m-M)= 31.24\pm 0.17$&(5)\\
\hline
Heliocentric velocity&$\varv_{\mathrm{helio}}= 906 \pm7\,\mathrm{km\,s}^{-1}$& (3)\\
Hubble type & E0+ & (4) \\
Ellipticity &$\epsilon = 0.15$ &(6)\\
Position angle  & $PA=145^\circ$& (5)  \\
Effective radius &${R_{\mathrm{e}}}=101\farcs\,7\,(=7.5\,\mathrm{kpc})$ & (7)\\
\hline
Blue magnitude & $B_{\rm{T}}=10.43 $ & (4)\\
Total apparent magnitude &T1$=8.70\pm 0.05$  & (5)\\
\hline
Metallicity&$[\mathrm{ Fe/H} ] = -0.01$\,dex & (7)\\
X-ray luminosity&$L_{\mathrm{X}}=1.78\pm 0.10\times\!10^{\,41}\,\textrm{ergs/s}$&  (8)\\
Nuclear X-ray emission & $\leq 2.7\times 10^{\,38}\, \textrm{ergs/s}$ &(9)\\
\hline
\hline
\end{tabular}
\end{center}
\end{table}
In the 1970s NGC\,4636 was the target of several
\ion{H}{i}-observations. Despite some claims of an extended
\ion{H}{i}-emission (Knapp et al.~\cite{1978AJ.....83...11K},
Bottinelli \& Gouguenheim~\cite{1978A&A....64L...3B}), the deepest
observations performed by  Krishna Kumar \& Thonnard
(\cite{1983AJ.....88..260K}), yielded an upper limit of only $11.5
\times 10^{7} {\rm M}_{\sun}$ to the total \ion{H}{i} mass of NGC\,4636.

Because of its high X-ray luminosity, NGC\,4636 has been studied
extensively in this frequency range and these observations have also
been used to infer its dark matter content and distribution.

Forman et al.~(\cite{1985ApJ...293..102F}) analyzed Einstein X-ray
data from a large sample of early type galaxies. For NGC\,4636 they found
extended emission (out to a radius of
${R_{\mathrm{max}}}=44\,\textrm{kpc}$) and derived an X-ray luminosity
of ${L_{\mathrm{X}}}=1.78\pm 0.10\times 10^{41} \textrm{ergs s}^{-1}$.
They quote a mass-to-light ratio $M_{\mathrm{total}}/L_{\mathrm{B}} =
87$ for $R_{\mathrm{max}}$, the second largest value in their sample.\\
Matsushita et al.~(\cite{1998ApJ...499L..13M}), using ASCA, found very
extended X-ray emission around NGC\,4636 out to 60\arcmin.  This
corresponds to a radius of almost 300\,kpc in a distance of
15\,Mpc. The authors suggest that the gravitational halo of NGC\,4636
itself terminates in the region between 10 and 30\,kpc.  They observe
an increase in mass beyond 30\,kpc forming a halo-in-halo structure,
whose size is comparable to that of a galaxy group.\\
Jones et al.~(\cite{2002ApJ...567L.115J}) used the Chandra X-Ray
Observatory Advanced CCD Imaging Spectrometer (ACIS) to study the
X-ray halo of NGC\,4636 with a spatial resolution of
$\sim\!50$\,pc. The observed X-ray features are so far unique: the
images show symmetric, arm-like structures which surround a bright
central region. These structures extend $\sim\!8\,\textrm{kpc}$ from
the galaxy center. The leading edges of these arms show changes in
brightness by a factor of two on scales of just a few
arcseconds. While these sharp fronts might be suggestive of a merger,
the authors favor a nuclear outburst as the source of the observed
morphology. They interpret the arms as the projected edges of
paraboloidal shock fronts expanding from the nucleus. Jones et
al.~speculate that the unusually high X-ray luminosity of NGC\,4636
might be due to the advanced state of the cooling that brought on the
outburst and may be further enhanced by the shocks. When the gas halo
returns to hydrostatic equilibrium, the X-ray luminosity could decline
by an order of magnitude or more.\\
Loewenstein~\& Mushotzky (\cite{loewenstein03}) used the Chandra X-Ray
Observatory to measure the hot interstellar medium temperature profile
of NGC\,4636. Assuming hydrostatic equilibrium, they determined the
total enclosed mass profile for the radial range $0.7 < r <
35\,\textrm{kpc}$.  These authors come to the conclusion that dark
matter constitutes a large fraction (between 50\% and 80\%) of the
total mass even inside the effective radius.  \\
Given all these
peculiarities, it is of great interest to use globular clusters as
dynamical tracers in order to evaluate the dark matter content of
NGC\,4636.  This work continues our programme on the dynamics of
globular cluster systems around early-type galaxies. Previous papers
on NGC\,1399, the central galaxy of the Fornax cluster are Richtler et
al.~(\cite{1399.2}) and Dirsch et al.~(\cite{1399.1}).
 \subsection{The NGC\,4636 Globular Cluster System}
\label{sect:clus}
The globular cluster system  of NGC\,4636 has been the target of three
photometric studies.
Hanes
(\cite{1977MmRAS..84...45H}) used photographic plates to assess the
number and the luminosity of the GCs. 
He derived a specific frequency (i.e.~the number of GCs normalized to
the host galaxy's luminosity (Harris\,\&\,van\,den\,Bergh
\cite{harris81}) of $S_{\mathrm{N}} = 9$, which is unusually high for
a quite isolated elliptical galaxy. Most field ellipticals have
$S_{\mathrm{N}}$ values between $3$ and $4$.  The first CCD-study was
performed by Kissler et al.~(\cite{1994A&A...287..463K}) who observed
a field of $7\farcm 5 \times 7\farcm 5$ centered on NGC\,4636 in one
band (Cousins $V$, limiting magnitude of $24.25$). They estimated the
total number of GCs to $3600 \pm 200$, which corresponds to
$S_{\mathrm{N}} = 7.5\pm 2.0$, hence supporting the unusually high
number of GCs.\\
The full extent of the GCS together with color information only became
apparent in the recent study of Dirsch et al.~(\cite{boris}). They
used the MOSAIC camera at the 4-m telescope at the Cerro Tololo
Interamerican Observatory (CTIO). The bands were Kron-Cousins R and
Washington C. The final images have a field-of-view of $34\farcm 7
\times 34\farcm 7$. The color distribution of the GCs shows the
familiar bimodal shape. The color of the minimum of the inner sample
(C-R=1.55) was chosen to divide red (metal-rich) from blue
(metal-poor) clusters, frequently interpreted as two different
sub-populations (e.g.~Ashman~\& Zepf \cite{1998gcs..book.....A}, Kundu
\& Whitmore \cite{kundu}, Larsen et al.~\cite{larsen}) .  For the
total number of globular clusters Dirsch et al.~derived $3700\pm 200$,
which agrees well with the previous measurement. The specific
frequency $S_{\mathrm{N}}$ ranges between $5.8\pm 1.2$ and $8.9\pm
1.2$, for the GCLF and SBF distance, respectively.  An interesting
feature shows up in the radial distribution of the clusters: While the
(projected) radial number density distribution of the blue clusters
can be described by one power-law ($\mathrm{n}(r)\propto r^\alpha$)
within 1\arcmin - 8\arcmin, the density distribution of the red
clusters changes the exponent within this radial interval. 
At a radial distance of 7\arcmin\, (9\arcmin\,) for the red (blue)
clusters, the power-law exponent drops to about $-5$ (cf.~Table\,3 in
Dirsch et al.~\cite{boris}) Such behavior has not been observed in any
GCS before. This abrupt decrease in number density might mark the
limit of the GCS and possibly that of the galaxy itself. The
photometric data by Dirsch et al.~forms the basis of the candidate
selection for our spectroscopic observations described in the
following section.
\section{Observations}
The observations were carried out during four nights (May 7--10 2002)
at the European Southern Observatory (ESO) Very Large Telescope (VLT)
facility at Cerro Paranal, Chile. The VLT Unit Telescope\,4 was used
with the FORS2 (FOcal Reducer/low dispersion Spectrograph) instrument
equipped with the Mask EXchange Unit (MXU). The standard resolution
collimator used for this program provided a field-of-view of $6\farcm
8 \times 6\farcm 8$.  The detector system consisted of two
$4096\times2048$ red optimized CCDs with a pixel size of
$15\mu\textrm{m}$. For our observations the chips were read out using
a $2\times2$ binning. The grism 600B gave a spectral resolution of
about 3\,\AA. The spectral coverage was dependent on the slit position
on the mask. In most cases, the usable coverage was about 2000\,\AA\ with
limits on the red side varying between 5500 and 6500\,\AA.  We exposed
13 spectroscopic masks, the preparation of which is described in the
next section.  Flat fielding was done with internal flat lamps. A
Hg-Cd-He lamp was used for wavelength calibration. The observations
are summarized in Table\,\ref{ov}.
\subsection{Mask Preparation}
\begin{figure}[t]
\centerline{\resizebox{\hsize}{!}{\includegraphics[angle=-90]{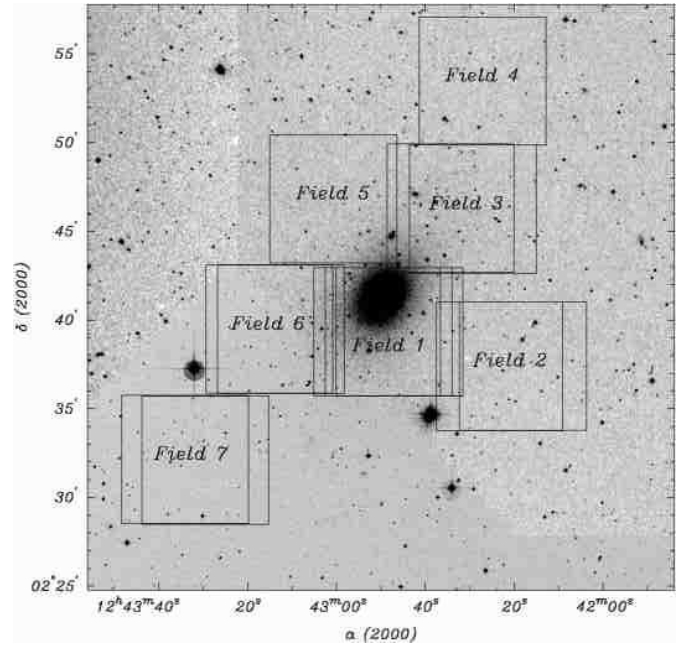}}} 
\caption[Positions of the fields]{{Positions of the fields overlaid on
    a $33\arcmin \times 33\arcmin$ DSS image centered on
    NGC\,4636. For each mask the entire FORS\,2 $6\farcm 8 \times
    6\farcm 8$ field-of-view is shown. Note that the slit positions
    are confined to the central $4\farcm 3 \times 6\farcm 8$ 
region of each pointing.}}
\label{fieldsdss}
\end{figure}

Pre-images of the 13 pointings in 7 fields (see Fig.~\ref{fieldsdss})
were obtained in service mode in October 2000.  Each pointing was
observed in Cousins V and I filters for 30 seconds. The GC candidate
selection was based upon the photometric work presented in Dirsch et
al.~(\cite{boris}). Cluster candidates had to fulfill the following
criteria: they had to be brighter than 22.7 in $R$ and the allowed
color range was $0.9 < C-R < 2.1$. Further, the candidates should
exhibit a star-like appearance on the pre-images to distinguish them
from background galaxies.  The ESO FORS Instrumental Mask Simulator
(FIMS) software was then used to select the positions, widths and
lengths of the slits. A slit width of 1\arcsec\, was chosen according
to the sub-arcsecond seeing conditions on Paranal.  To maximize the
number of GC spectra obtained per mask, we chose to observe our
targets and sky positions through separate slits of 2\arcsec\,
length. Obviously, the quality of the sky subtraction then depends on
the accuracy of the wavelength calibration.  As it will be shown, it
turned out to be satisfactory to obtain radial velocities with the
desired accuracy. After positioning the slits for the selected GC
candidates, the remaining space on the masks (especially in the outer
fields) was used to include additional objects.  Thus, also background
galaxies and point sources not matching the above-mentioned criteria
were observed.
\subsection{The Data Set}
The MXU-observations of our thirteen fields have exposure times in the
range of 3600 to 7200 seconds (see Table\,\ref{ov} for a summary of the
observations). To minimize the effects from cosmic ray hits,
the observation of each mask was divided into two exposures - with the
exception of Field 1.1 for which three science images were
obtained. In addition to the spectroscopic observations, calibration
measurements, i.e.~bias, flat fields and wavelength calibration
exposures were obtained during day time.
\begin{table*}
\centering
\begin{tabular}{cllccrcrl@{-}l@{-}lr@{:}l}\hline
\hline
{Field} & \multicolumn{2}{c}{Center Position} &
{Exp.~Time}&{Seeing}&\multicolumn{2}{c}{\#\,Slits}&
\multicolumn{1}{c}{\#\,GCs}&
\multicolumn{3}{c}{Night}&\multicolumn{2}{c}{UT} \\
{} & \multicolumn{2}{c}{{({J\,2000})}} &
{{({sec})}}&{{{}}}&\multicolumn{1}{c}{{total}}&
\multicolumn{1}{c}{{objects}}&
\multicolumn{1}{c}{\,}&
\multicolumn{3}{c}{\,}&
\multicolumn{2}{c}{(start)\,} \\
\hline
1.1& 12:42:48.2& 02:39:21.0 & 7200  & 1 \farcs 1& 111&79  & 39 &2002 &05&07& \ 23&27 \\
1.2& 12:42:45.8& 02:39:24.2  & 4500  & 0 \farcs 9&98 &61  & 47 &2002&05&07& 01&49 \\
1.3& 12:42:50.8& 02:39:21.3 & 5400  & 1 \farcs 2& 106&65  & 26 &2002&05&07&
03&17  \rule[-1.5ex]{0pt}{2ex} \\
2.1& 12:42:17.8& 02:37:27.9& 3600 & 1 \farcs 0 & 101& 51  & 3&2002&05&07&05&11 \\
2.2& 12:42:23.0& 02:37:27.6 & 3600 & 1 \farcs 0& 99& 48  & 2&2002&05&08&23&32  \rule[-1.5ex]{0pt}{2ex} \\
3.1& 12:42:34.2& 02:46:22.4 & 3600 & 0 \farcs 8&113 &77 & 26 &2002&05&08 &01&21 \\
3.2& 12:42:28.9& 02:46:20.7 & 3600 & 1 \farcs 1&113 &54 & 14 &2002&05&08&02&31
\rule[-1.5ex]{0pt}{2ex} \\
4.1& 12:42:27.1& 02:53:27.4 & 3600 & 1 \farcs 3&98 &44 & 2& 2002&05&10&23&37
\rule[-1.5ex]{0pt}{2ex} \\
5.1& 12:43:00.6& 02:46:50.3 & 3600 & 0 \farcs 9 &105 & 17 & 59&2002&05&08&
03&43  \rule[-1.5ex]{0pt}{2ex} \\

6.1& 12:43:15.0& 02:39:33.7& 3600 & 1 \farcs 2& 107 &53 & 8&2002&05&10&01&04 \\
6.2& 12:43:12.6& 02:39:32.0& 3600 & 0 \farcs 9& 110 & 55 & 16 &2002&05&09&00&18  \rule[-1.5ex]{0pt}{2ex} \\
7.1& 12:43:33.9& 02:32:12.1 & 3600 & 0 \farcs 9&109 & 52& 0&2002&05&09&01&31 \\
7.2& 12:43:29.3& 02:32:09.7& 3600 &  0 \farcs 8&113 & 59& 2&2002&05&09&02&44 \\
\hline
\hline

\end{tabular}
\caption[Summary of observations]{{Summary of observations (ESO
    program ID 69.B-0366(B)). The seeing values were determined from the acquisition images taken just before the MXU exposures.}}

\label{ov}
\end{table*}
\section{Data Reduction}
The following section summarizes the reduction steps performed to
obtain the final sky-subtracted spectra. Most of the data reduction
was done within the {IRAF}-environment\footnote{IRAF (Image
Reduction and Analysis Facility) is distributed by the National
Optical Astronomy Observatories, which are operated by the Association
of Universities for Research in Astronomy, Inc., under cooperative
agreement with the National Science
Foundation. (http://iraf.noao.edu/)}.
\subsection{Basic Reduction}
The merging of the two CCD-frames was done with \emph{fsmosaic} which
is available as a part of the ESO-FIMS software. The dimension of the
combined images is $2050\times 2076$ pixels.  For cosmic ray removal,
the task \emph{bclean} from the \emph{Starlink Figaro}-package was
found to perform best. Biasing and flat-fielding was done in
the standard way.  Since the aim of this study is the measurement of
radial velocities rather than the measurement of line strengths, we
did not apply any response function fitting or flux calibration.  To
obtain one-dimensional spectra (aperture definition, tracing, and
extraction) the IRAF task \emph{apall} from the
\emph{apextract} package was used.
Once the aperture positions were defined on the science images, they
were checked and - where necessary - adjusted on the flatfield divided
exposures. 
For the 2\arcsec\, long slits used for most sources, an aperture size of
4\,pixels (= 1\arcsec) was found to yield the best results.
\subsection{Wavelength Calibration and Sky Subtraction}
The one-dimensional arc lamp spectra were calibrated using the
interactive task \emph{identify} from the \emph{noao.onedspec}
package.
The dispersion/wave\-length solution was approximated by a
seventh-order Chebyshev polynomial. The residuals of this fit were
inspected for outliers, and if necessary the deviant data points were
excluded from the fit in the next step. The RMS errors of the final
fits were about 0.04 \AA. 
As described above, sky spectra were measured through separate
slits. In order to perform the sky subtraction for a given
GC-spectrum, two or three adjacent (within 18\arcsec) of these sky
spectra were combined. This combined spectrum was then subtracted
using the \emph{skytweak} task.
\section{Measurement of radial velocities}
The {IRAF} task \emph{fxcor} from the \emph{noao.rv} package performs
a Fourier cross-cor\-re\-la\-tion on the input object and template
spectra. This program is based upon the technique developed by
Tonry~\& Davis (\cite{1979AJ.....84.1511T}).  For the
cross-correlation, a template with a high S/N and a spectrum similar
to that of a globular cluster is required. NGC\,4636 itself, however
is not suitable due to its high velocity dispersion of about
200\,$\textrm{km\,s}^{-1}$. Therefore, the velocities were measured
using a spectrum (S/N $\sim\!30$) of NGC\,1396 as template. This
spectrum was obtained in Dec.~2000, during observations of the
NGC\,1399 GCS, using the same FORS2 instrumental setup and has already
been used as template for investigating the kinematics of the
NGC\,1399 GCS (Dirsch et al.~\cite{1399.3}). NGC\,1396 is a dwarf
elliptical galaxy close to NGC\,1399 , and its spectrum resembles that
of a metal-rich globular cluster. Its velocity dispersion is only
$65\pm 6 \textrm{km\,s}^{-1}$ (Wegner et al.~\cite{wegner04}). We
adopted a heliocentric radial velocity of $815\pm
8\,\mathrm{km\,s^{-1}}$ (Dirsch et al.~\cite{1399.3}) .  The
velocities obtained with this template were checked with a second
template spectrum obtained during our observations (object 1.3:56), a
bright star whose observed velocity was found to be
-682\,$\mathrm{km\,s}^{-1}$\, with respect to the galaxy template.
The velocities measured with the two different templates agree very
well. Since the \emph{fxcor} uncertainty estimates were smaller for
the galaxy template, only these velocities are used for the subsequent
analysis.  The accuracy of the velocity determination depends on the
shape of the spectrum, and the wavelength range for which the
cross-correlation is performed. A range of $4700< \lambda <
5500\,\textrm{\AA}$ was found to be an appropriate choice in most
cases. The upper limit was chosen to avoid the night sky emission
lines found around the strong [OI] 5577\,\AA\, feature. In some
spectra however, the relatively weak Nitrogen line at 5199\,\AA\, left
substantial residuals: where this was the case, the wavelength
interval was adjusted accordingly.  For the velocities of the
NGC\,4636 GCS, we applied a heliocentric correction of
$\varv_{\mathrm{hc}}=-18\,\mathrm{km\,s}^{-1}$; the correction for the
NGC\,1396 spectrum is $\varv_{\mathrm{hc}}=-10\,\mathrm{km\,s}^{-1}$.
The uncertainty for the velocities is estimated using:
$$     \Delta \varv_{\textrm{}} = \sqrt{ \Delta \varv^2_{\mathrm{fxcor}}+\Delta
       \varv^2_{\mathrm{template}}}\;,
$$
where $\Delta \varv_{\mathrm{fxcor}}$ is the velocity uncertainty output
by \emph{fxcor}, and $\Delta \varv_{\mathrm{template}}$ is the velocity
uncertainty of the template spectrum.
\subsection{Zero Point Shifts}
Since the wavelength calibration exposures for each mask were obtained
during the day following the observation, the mask was moved into the
focal plane again. The finite positioning accuracy of the MXU, and
shifts due to instrument flexure under gravity - all calibration
exposures are obtained with the telescope pointing at the zenith -
result in an offset between the science and the calibration
images. This offset manifests itself as a zero point shift in the
observed wavelength of the strong telluric [OI] 5577\AA\, emission
line.
The zero point shifts in our data set were found to vary with the
position of the slit along the $y$-axis of the mask.
To quantify this effect, the wavelength of the 5577\AA\, line was
measured in all spectra for a given mask. A linear function
was fit to the data and provided 
a velocity correction for every aperture.  These
corrections have values in the range  $
-25\textrm{$\mathrm{km\,s}^{-1}$}\leq\Delta \varv_{\mathrm{MXU}}\leq 25\,
\textrm{$\mathrm{km\,s}^{-1}$}$.
\label{sect:v4636}

To measure the systemic velocity of NGC\,4636, we used the nine
spectra of NGC\,4636 from mask 1\_1, and found a heliocentric velocity
of $906\pm 7\,\textrm{$\mathrm{km\,s}^{-1}$}$.  In the NASA/IPAC
Extragalactic Database (NED)\footnote{The NASA/IPAC Extragalactic
Database (NED) is operated by the Jet Propulsion Laboratory,
California Institute of Technology, under contract with the National
Aeronautics and Space Administration.} one finds seven optical
velocity measurements with uncertainty estimates. The mean is $917\, \pm
31\textrm{$\mathrm{km\,s}^{-1}$}$. Hence, our velocity agrees very
well with previous measurements.
\subsection{Velocity Uncertainties}
\begin{figure}
\centering
{\resizebox{0.45\textwidth}{!}{\includegraphics[angle=0]{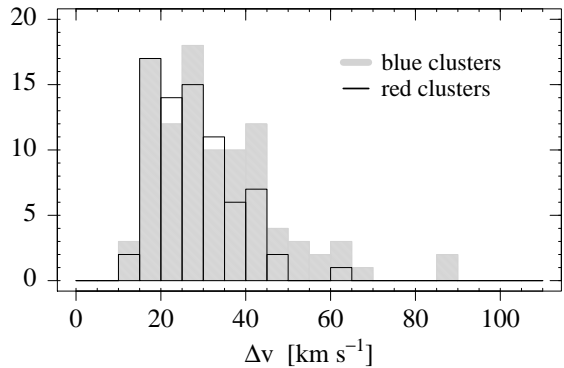}}}
\caption[]{ Histogram of the velocity uncertainties calculated by the
\emph{fxcor}--task.  The grey histogram shows the values for the blue
clusters, the black histogram shows the same for the red clusters.}
\label{fxerr}
 \end{figure}

\begin{figure}
\centering
{\resizebox{0.45\textwidth}{!}{\includegraphics[angle=0]{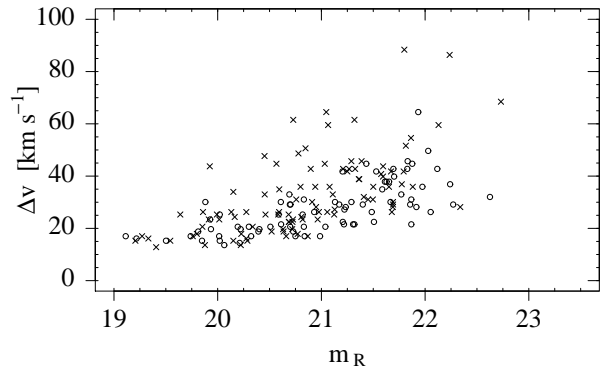}}}
\caption[]{{Velocity uncertainties versus R-magnitude for the globular
    clusters (crosses: blue clusters; open circles: red clusters).}}
\label{fxmag}
\end{figure}

For the GCs, no trend is visible for the metal poor (blue) and the
metal rich (red) subpopulations: both have very similar error
distributions (see Fig.~\ref{fxerr}).  Figure\,\ref{fxmag} shows the
velocity uncertainties as a function of apparent R-magnitude. As
expected, the measurements of fainter objects tend to have larger
uncertainties.  The uncertainties returned by the \emph{fxcor} task
are computed according to the formulae given by Tonry and Davis
(\cite{1979AJ.....84.1511T}).  To check whether these uncertainties
are a realistic error estimate, we compared the velocities of those 40
objects for which we had two spectra since they were present on two
masks.The mean difference for the velocity measurements was found to
be very similar to the mean of the uncertainties given by
\emph{fxcor}. Hence, the \emph{fxcor} uncertainties can be considered
a good error estimate.  The velocities determined for the GCs are listed
in appendix \ref{sect:GCs} which is available in electronic form.
\section{Properties of the Globular Cluster Sample}    
For every object with a velocity measurement, the color and magnitude
information was extracted from the photometry-database from Dirsch et
al.~(\cite{boris}) via coordinate matching. The coordinates,
velocities, $\mathrm{C-R}$ -colors and apparent R magnitudes of these
objects are listed in appendix \ref{sect:GCs}. For two clusters, neither
color nor magnitude information exists, since they happened to be
located near a bad row of the MOSAIC-CCD.
\subsection{Defining the Sample}
\begin{figure}
\centerline{\resizebox{0.5\textwidth}{!}{\includegraphics[angle=0]{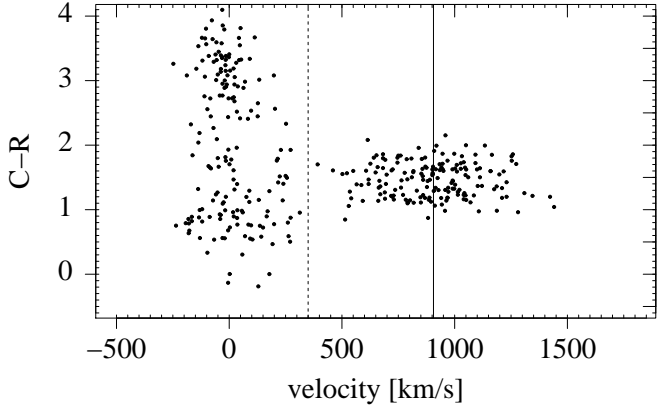}}}
\caption[]{
  Color vs.~radial velocity.  All objects with heliocentric velocities
  higher than $350\,\textrm{km\,s}^{-1}$ \,(dotted line) are regarded
  as globular clusters. The objects with velocities falling below this
  limit are foreground stars.  The solid line marks the systemic
  velocity of NGC\,4636.}
\label{gcstar}
\end{figure}

The first step was to divide our data set into GCs and foreground
stars. A compelling distinction between stars and globular clusters
would be given if the GCs could be resolved on the acquisition
images. However, only very few, about three, of our clusters are
marginally resolved. 
Guided by Fig.~\ref{gcstar} which shows the distribution of velocity
versus color for all measured objects, a lower limit of
$350\,\textrm{km\,s}^{-1}$ for the radial velocity of the GCs was
chosen.  However, one has to be aware that there might be a velocity
domain where this identification is doubtful.  Given that the Galactic
rotation vector towards the position of NGC\,4636 is
$-78\,\textrm{km\,s}^{-1}$, one cannot exclude that foreground stars
have heliocentric velocities of up to
$400\,\textrm{km\,s}^{-1}$. Neither can one rule out the presence of
GCs with radial velocities of more than $500\,\textrm{km\,s}^{-1}$
relative to NGC\,4636.
 
\begin{figure}[]
\centering
\centerline{\resizebox{\hsize}{!}{\includegraphics[angle=0]{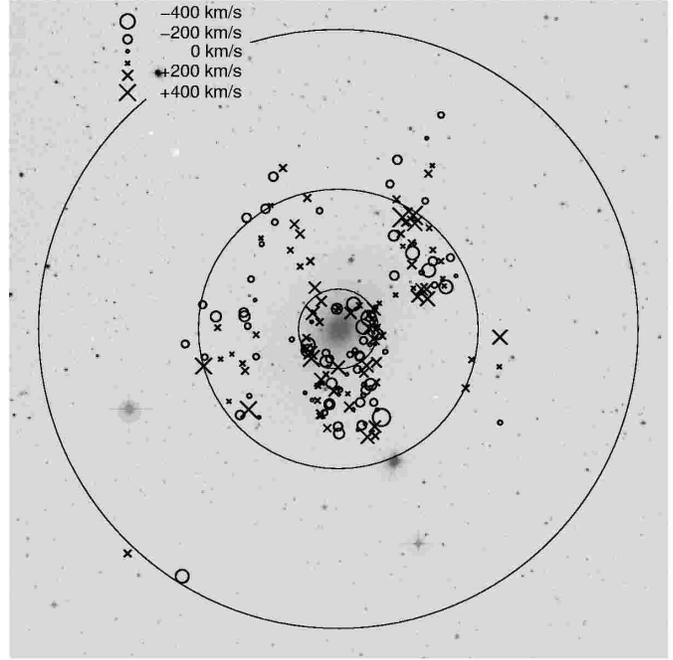}}}
\caption[]{Spatial distribution of the globular clusters overlaid on
    a $33\arcmin \times 33\arcmin$ DSS-image centered on
    NGC\,4636. North is to the top, East to the left. GCs with
    velocities larger than the systemic velocity of NGC\,4636 are
    marked by crosses. Circles denote GCs with negative relative
    velocities.The concentric circles have radii of 2, 7, and 15
    arcminutes) }
\label{spatdss}
\end{figure}

The range in velocities occupied by foreground stars and globular
clusters is $-248\leq \varv\leq313\,\textrm{km\,s}^{-1}$ and $392\leq
\varv\leq1441\,\textrm{km\,s}^{-1}$, respectively.  The data set
comprises 174 globular clusters and 175 foreground stars. The basic
properties of the GC sample are listed in Table\,\ref{gcsample}. The
so-defined GCs appear in Fig.~\ref{spatdss} which shows their
distribution around NGC\,4636. The inner region is well sampled, but
due to the sharp drop-off at $\sim\!7\arcmin$ only a handful of
clusters is located outside this radius.
\begin{table*}
\centering
\begin{tabular}{llll}\hline\hline
&{\bf{all}} &{\bf{blue}} &{\bf{red }} \rule[-0.5ex]{0ex}{3ex}\\ \hline
number of GCs & 174 &  97  & 75 \\\hline
mean velocity $[\textrm{km\,s}^{-1}] $& 904 & 911  &
887\\
min. velocity $[\textrm{km\,s}^{-1}]$ & 392 & 514
& 392  \\
max. velocity $[\textrm{km\,s}^{-1}]$ & 1441 &
1441 & 1273 \\
mean vel. uncertainty $[\textrm{km\,s}^{-1}] $& 31 &
33  & 28 \\
velocity dispersion $[\sigma(R)\,\textrm{km\,s}^{-1}] $ & $203\pm11$
& $202\pm15$ & $199\pm17$ \\
\hline
color range C-R &   & $0.85$ -  $1.55$  & $1.55$ - $2.15$ \\
range in projected radius& 0\farcm 91 - 15\farcm 43& 0\farcm 91 -
15\farcm 43 &1\farcm 00 - 11\farcm 90  \\
\hline\hline
\normalsize
\end{tabular}
\caption[Basic properties of the GC sample]{{Basic properties of the
    globular cluster sample}}
\label{gcsample}
\end{table*}
\subsection{Color and Luminosity Distribution}
\label{sect:col}
\begin{figure}
\centering

{\resizebox{0.46\textwidth}{!}{\includegraphics[angle=0]{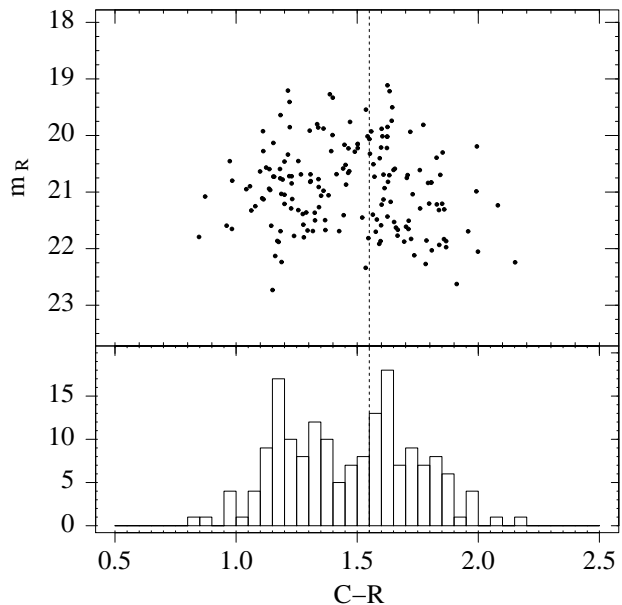}}}
\caption[]{{{\bf{Top}:} Color magnitude diagram for the globular
    clusters. Plotted are C-R\, vs.~R. {\bf{Bottom}:}\, Color
    distribution for the GCs (bin width = 0.05\,mag). In both panels,
    the dotted line at $\rm{C-R} = 1.55$ indicates the color dividing
    blue from red clusters (Dirsch et al.~\cite{boris}) .

}}
\label{cmd}
\end{figure}

\begin{figure}
\centering
{\resizebox{0.45\textwidth}{!}{\includegraphics[angle=0]{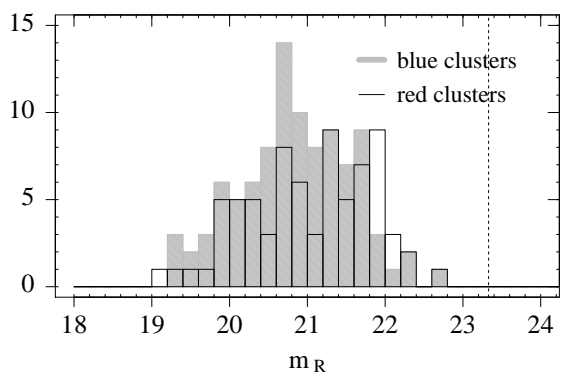}}}
\caption[]{{Globular cluster luminosity
    distribution (bin width = 0.2\,mag). The dotted line at 23.33
    indicates the turn-over magnitude of the GCS. 
    Histograms for blue
    (solid grey) and red  clusters.}}
\label{gclf}
\end{figure}

For any further investigation,
it is important that the color distribution of the GCs with measured
radial velocities resembles that of the entire cluster system. Indeed,
the bimodality of the color distribution found in the photometric
study by Dirsch et al.~(\cite{boris}) is clearly visible in our much
smaller sample (see Fig.~\ref{cmd}). For consistency,
the same dividing color of $\mathrm{C-R} = 1.55$ was used. Thus, we find
$97$ blue and $75$ red globular clusters. The color magnitude diagram
(CMD) for the GCs is shown in Fig.~\ref{cmd}.
Although the selection of GC candidates was obviously biased towards
bright clusters, no clusters brighter than $R=19$ were found. 
A few clusters are fainter than $R=22$, but
their velocities have large errors (cf.~Fig.~\ref{fxmag}) and should
be considered with caution.
\label{sect:lum}
Figure~\ref{gclf}  shows the histogram of the apparent R-mag\-ni\-tudes
for the globular clusters. It  illustrates that our
spectroscopic study only probes the very bright part of the GCLF, well
above the turn--over magnitude at 23.33 (Dirsch et al.~\cite{boris}).
{{Furthermore, it shows that there is no significant
difference between the distributions of blue and red clusters}}.  The
magnitude range of the GCs is $19.11 \geq m_{\,\mathrm{R}} \geq 22.73$
in Kron R.  When converted to absolute magnitudes (using the distance
modulus of 30.83, Tonry et al.~\cite{2001ApJ...546..681T}), the
globular clusters in our sample span the range $-11.6\leq
M_{\mathrm{R}} \leq -8.1$ in R and $-11.25\leq M_{\mathrm{V}} \leq
-7.6$ in $V$ (for $V\!-R = 0.47$, Dirsch et
al.~\cite{boris}). Assuming a typical mass-to-light ratio of
$M/L_{\mathrm{V}} =2$, one finds that the masses of the brightest
clusters are of the order $7-8\times 10^{ 6}$ ${\rm M}_{\sun}$.  For
comparison: the absolute visual magnitude of $\omega$ Centauri
(NGC\,5139), is $M_{\mathrm{V}} = -10.29 $
(Harris~\cite{1996AJ....112.1487H}) and its mass is about $5\times 10^{\,6}\,{\rm M}_{\sun}$.
\subsection{Spatial Distribution}
\label{sect:spatial}
As can be seen in Fig.~\ref{spatdss}, our target clusters are not
uniformly distributed with respect to the center of NGC\,4636. At the
time of the mask preparation - before the photometric data set was
fully analyzed - the projected globular cluster number density 
was overestimated.  Some fields
were placed at larger projected distances (see Fig.~\ref{fieldsdss}) -
at the expense of a more uniform azimuthal coverage. Yet in total, only
15 of the 174 GCs in our sample are found at projected radii larger
than $7\farcm 5$. Thus, any measurement of the line-of-sight velocity
dispersion beyond this radius is clearly doubtful.  Note that the
radius at which the power-law index of the projected globular cluster
density drops (cf.~section~\ref{sect:clus}) lies in the range 7\arcmin
to 9\arcmin. 
\subsection{Velocity Distributions}
\begin{figure}[]
\centering
{\resizebox{0.45\textwidth}{!}{\includegraphics[angle=0]{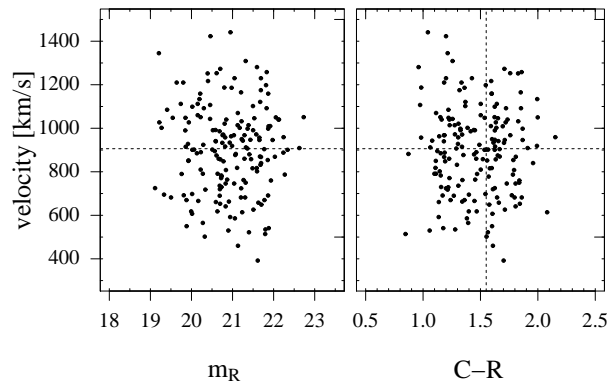}}}
\caption[]{{{
      {\bf{Left}:\,} Radial velocity as a function of apparent
      R-magnitude.}}{ {\bf{Right}:\,} Radial velocity of the GCs
    plotted vs.~the C-R -color.}}
\label{fig:9}
\end{figure}

\begin{figure*}
\centering
      {\includegraphics[width=\textwidth,angle=0]{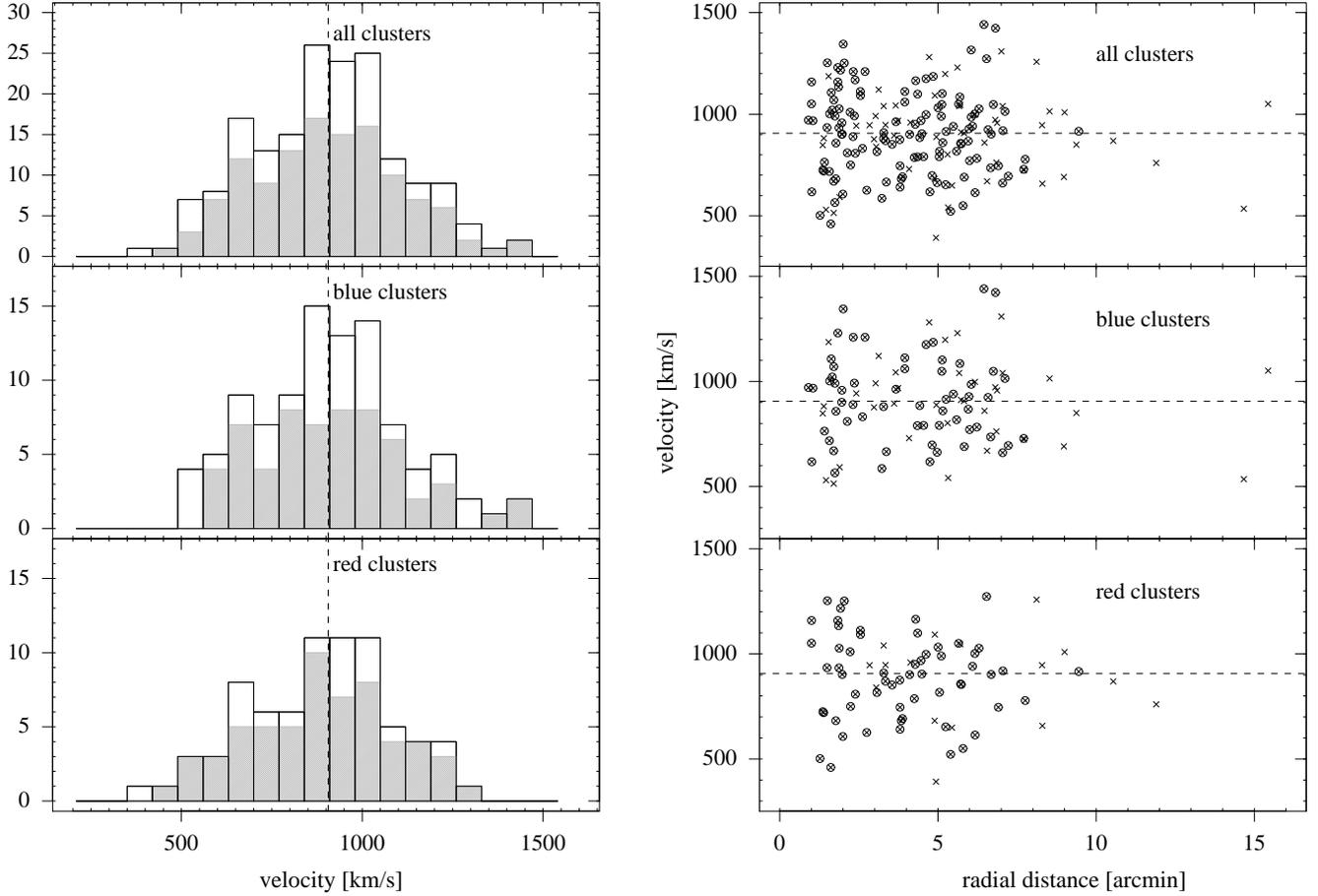}}
\caption{{ The left panels show the histograms of the GC velocities
for the entire, the blue and the red subsample, respectively. In each
panel, the grey histogram shows the distribution of the corresponding
error-selected ($\Delta \varv < 35\,\textrm{km\,s}^{-1}$) sample. The
dashed line indicates the systemic velocity of NGC\,4636, and the bin
width is $70\,\textrm{km\,s}^{-1}$. The right panels show the radial
velocities versus projected galactocentric distance for all, the blue
and the red clusters. The crosses in circles denote data points with
$\Delta \varv < 35\,\textrm{km\,s}^{-1}$, i.e.~they form the error
selected samples. Plain crosses denote data points with larger
velocity uncertainties. Again, the dashed line marks the systemic
velocity of NGC\,4636.} }
\label{fig10}
\end{figure*}

In the left panel of Fig.~\ref{fig:9} the velocities are  plotted 
against the  R magnitudes.
As expected, they are found to be concentrated towards the systemic
velocity. The brightest clusters however, tend to avoid the systemic
velocity. A larger sample is needed to confirm whether this effect is
real or due to low-number statistics.
The right panel of Fig.~\ref{fig:9} shows the velocities plotted
against the C-R\, colors of the clusters. Again, no clear trend is
visible, although one notes that the five clusters with the
highest velocities are blue.
The top right panel of Fig.~\ref{fig10} plots all GC radial velocities
versus the projected radii in arcminutes. This figure once more shows
how few velocities were obtained for large radii. The more important
observation, however, is that the scatter of the velocities about the
systemic velocity is rather small at a radial distance of about
2\arcmin -4\arcmin. As will be shown later (see section
\ref{sect:dispersion}), this translates directly into a decrease of the
projected velocity dispersion in this radial interval. On the other
hand, only relatively few GCs fall into this radial range.
The middle and lower right panels of Fig.~\ref{fig10} show the same
diagram for blue and red clusters, respectively. In the radial range
around 2\arcmin-4\arcmin, the scatter about the systemic velocity is
larger for the blue than for the red clusters, hence the radial
behavior of the velocity dispersions for the subsamples will turn out
to be rather different (see section \ref{sect:dispersion}).

\section{Rotation}
\begin{figure}[]
\centering
{\resizebox{0.45\textwidth}{!}{\includegraphics[bb=179 34 562 537,clip=,angle=270]{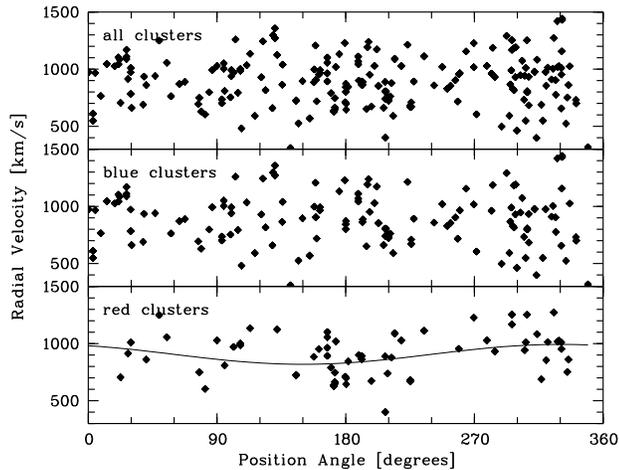}}}
\caption[]{This plot shows radial velocity
  vs. the position angle (North past East). The upper panel plots the
  entire sample, the middle panel the blue clusters and the lower
  panel the red clusters. A marginal rotational signal is detected in
  the entire sample, which is due to the stronger signal of the red
  clusters, which seem to rotate around the minor axis with an
  amplitude of about $90\,\textrm{km\,s}^{-1}$. The blue clusters do
  not rotate.}
\label{rotall}
\end{figure}

Due to the inhomogeneous azimuthal coverage of our sample it is
obvious that statements concerning rotation have to be considered with
caution. 
The diagnostic diagram which we have to analyze is a plot of radial
velocities vs. the position angle. C\^ot\'e et
al.~(\cite{2001ApJ...559..828C}) give a useful discussion of the
relation between the intrinsic and projected rotational velocity field
of a spherical system and we do not repeat that here.  If the
intrinsic rotation velocity field is stratified on spheres, and the
galaxy is not seen pole on, we measure radial velocities that depend
sinusoidally on the azimuth angle. Therefore, we fit the following
relation :
\begin{equation}
  \varv_r(\Theta)  = \varv_{sys} + A \sin(\Theta -\Theta_0), 
\end{equation} 
where $\varv_r$ is the measured radial velocity at the azimuth
angle $\Theta$, $\varv_{sys}$ is the systemic velocity, and $A$ the
rotation amplitude. Figure~\ref{rotall} shows the radial velocities of
three samples versus the azimuth angle which goes from North past
East. The uppermost panel is the full sample, then follow the blue and
the red clusters. A very marginal rotation signal is present in the
full sample with the parameters $A = 28 \pm 18\,\textrm{km\,s}^{-1}$
and $\Theta_0 = 63 \pm 43$. The blue sample does not show a
significant signal ($A = 11 \pm 27\,\textrm{km\,s}^{-1}$), while the
red sample shows significant rotation, which leaves its imprint on the
whole sample. The signal and its significance depends on the color
dividing blue and red clusters. The strongest signal is found for
clusters redder than C-R = 1.6. In this case, we find for the entire
red sample, $A = -87 \pm 18\,\textrm{km\,s}^{-1}$ and $\Theta_0 = 60
\pm 22$, which means a rotation around the minor axis with the
southeastern region approaching.  Both the amplitude and the position
angle remain within the uncertainty limits for subsamples selected in
radial bins. Selecting red clusters with projected distances below 5
arcmin, yields $A = - 105 \pm 36\,\textrm{km\,s}^{-1}$ and $\Theta_0 =
72 \pm 24$. Radii less than 4 arcmin give $A = - 114 \pm
44\,\textrm{km\,s}^{-1}$ and $\Theta_0 = 72 \pm 25$.  However, a
confirmation of the rotation of the red clusters is desirable given
the relatively low number statistics and the inhomogeneity regarding
radius and position angle. Attempts to interpret the rotation, for
example, as a dynamical sign for a merger event signs seem therefore
premature. In any case, it may be that rotation influences the
velocity dispersion of the red clusters, making their use as dynamical
tracers more complicated.

\section{Velocity Dispersion}
\begin{figure*}[t]
\centering
    {\includegraphics[width=\textwidth,angle=0]{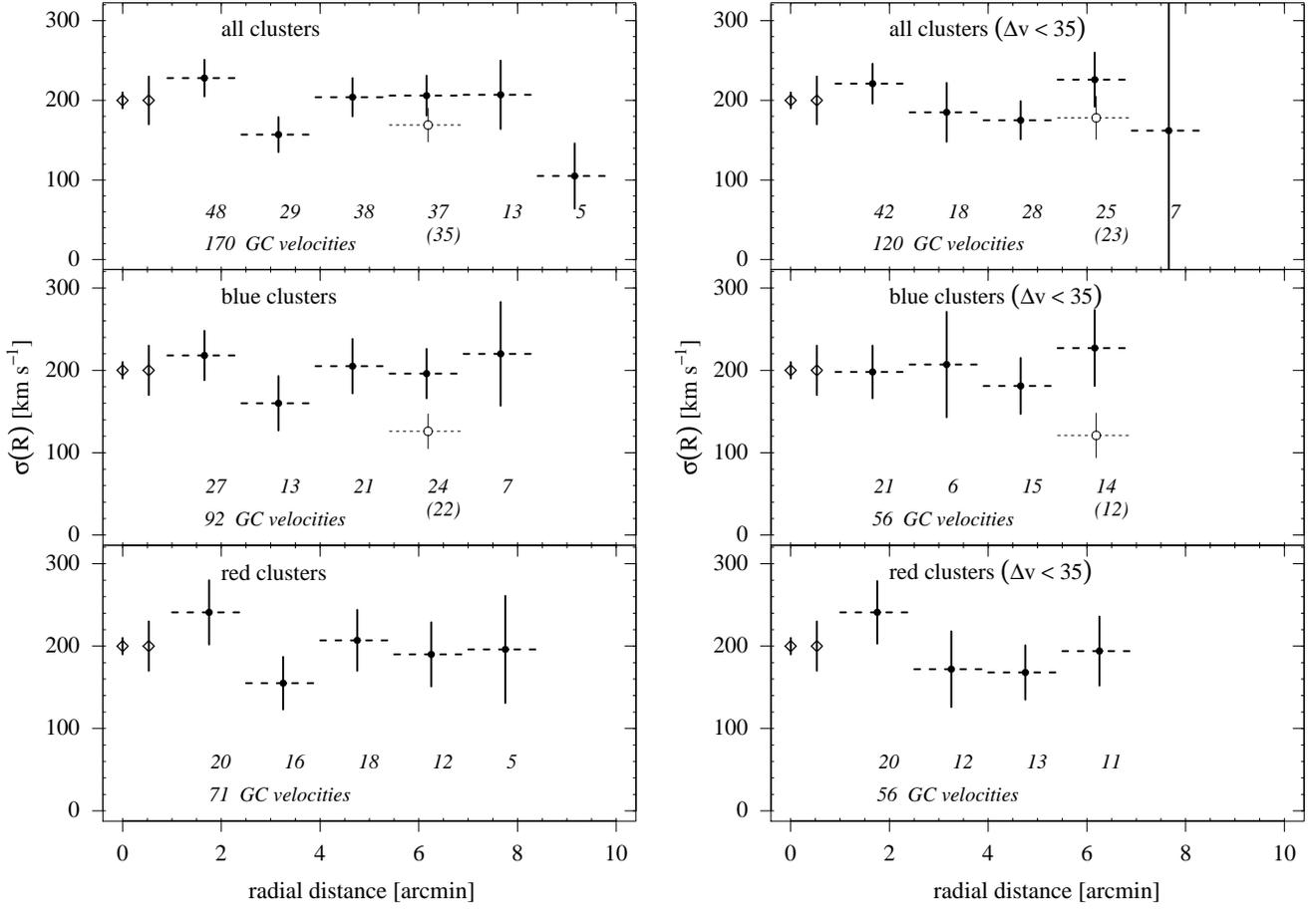}}
\caption{The left panels show the velocity dispersion versus projected
galactocentric radius for the whole, the blue and the red cluster
sample, respectively.  The right panels show the same for the
corresponding error-selected $\Delta\,\varv < 35\textrm{km\,s}^{-1}$
samples. The velocity dispersions were calculated for radial bins of
$1\arcmin 5$ width, and the number of velocities used for the
measurement is shown below each data point. The values plotted here
are listed in Table\,\ref{disptab}. In all panels, the diamonds
indicate the dispersion measurements from Bender et
al.~\cite{1994MNRAS.269..785B}. The open circles show the dispersion
values at 6\arcmin that are obtained when omitting the two (blue) GCs
with velocities above $1400\,\textrm{km\,s}^{-1}$.}
\label{dispall}
\end{figure*}

The top left panel of Fig.~\ref{fig10} shows the velocity distribution
of the entire velocity sample. The bin size was chosen to be
$70\,\textrm{km\,s}^{-1}$, which is larger than the mean
uncertainty. The mean velocity of $906\,\pm 16\,\textrm{km\,s}^{-1}$
is in excellent agreement with the systemic velocity of NGC\,4636 we
determined from the galaxy spectra (cf.~section \ref{sect:v4636}). The
shape of the distribution appears Gaussian (the Anderson--Darling test
for normality yields a $p$-value of 0.81 (Stephens
\cite{stephens})). This is what one would expect for a kinematical
sample that is close to isotropic and in dynamical equilibrium (see
e.g.~Binney~\& Tremaine \cite{BT}).
\label{sect:dispersion}
The most important observable which we want to extract from our data
is the projected velocity dispersion  and its dependence with radius. 
Throughout this analysis, it is assumed that our velocity
measurements are drawn from one normal distribution. 
To calculate the velocity dispersions presented in the following
section, we employed the maximum-likelihood dispersion estimator
presented by Pryor~\& Meylan (\cite{1993sdgc.proc..357P}). We fixed the
systemic velocity of NGC\,4636 to
$\varv_{\mathrm{sys}}=906\,\textrm{km\,s}^{-1}$.
Then, the velocity dispersion $\sigma$ is calculated (by iteration)
according to:
\begin{equation}
\sum \frac{(\varv_{i} - \varv_{\mathrm{sys}})^{\,2}}{{(\sigma^2 + {\delta_{i}^2})}^2} = \sum \frac{1}{\sigma^2 + {\delta_{i}^2}}\;\textrm{,}
\end{equation}
where the sum is taken over all velocities and the $\delta_{i}$ are
the uncertainties of the individual velocity measurements.  The
uncertainty of the resulting velocity dispersion is computed according
to the expression given by Pryor~\& Meylan.
Calculating the dispersion for all clusters yields:
$\sigma_{\mathrm{all}}  = 203\pm11\,\textrm{km\,s}^{-1}$.
The velocity dispersions for the blue and red cluster population  are
$\sigma_{\mathrm{blue}} = 202\pm15\,\textrm{km\,s}^{-1}$ and
$\sigma_{\mathrm{red}} = 199\pm17\,\textrm{km\,s}^{-1}$, respectively,
i.e.~they are equal within the uncertainties.
In order to study the radial dependence of the velocity dispersion,
we chose to divide our samples radial bins of $1\arcmin\!.5$ width.
The results are listed in Table\,\ref{disptab} and plotted in Fig.~\ref{dispall}.

\subsection{All Clusters}
{The upper left panel of Fig.~\ref{dispall} shows the radial
dependence of the velocity dispersion for all 170 GCs within $R\leq
9\farcm\,0$, For comparison, the line-of-sight velocity dispersion
measurements of the central region from Bender et
al.~(\cite{1994MNRAS.269..785B}) are shown as diamonds.  These authors
used a long-slit spectrum (with the slit positioned along the major
axis) to derive a central velocity dispersion of
$\sigma_{\mathrm{c}}=200\pm10\,\textrm{km\,s}^{-1}$ and for the
average velocity dispersion measured along the major axis ($R <
32\arcsec$), they found
$\sigma_{\mathrm{maj}}=200\pm30\,\textrm{km\,s}^{-1}$. Within the
uncertainties, our data points agree with the Bender et
al.~measurements.  Albeit consistent with a
constant velocity dispersion of about $200\,\textrm{km\,s}^{-1}$, this
plot suggests a decrease of $\sigma(R)$ by about
$50\,\textrm{km\,s}^{-1}$\, (from $\sim 220$ to $\sim
170\,\textrm{km\,s}^{-1}$) in the radial distance range of roughly
2\arcmin - 4\arcmin.  To check whether this behavior is merely an
artifact produced by velocity measurements with large uncertainties,
we next consider an error selected sample $(\Delta \varv <
35\,\textrm{km\,s}^{-1})$ (cf.~Fig.~\ref{fig10}, lower panel, to see
which data points are omitted). The plot for this smaller (120 GC
velocities) sample is shown in the upper right panel of
Fig.~\ref{dispall}.  Apparently, the decline is not due to large
individual uncertainties.  As will be shown next, this pattern is the
result of the different behavior of the blue and red
subpopulations. 
\subsection{Blue and Red Clusters}
The middle and lower panels of Fig.~\ref{dispall} show the radial
dependence of the velocity dispersion for the blue and the red
subpopulations, respectively. The values derived for the error
selected ($\Delta\,\varv < 35 \textrm{km\,s}^{-1}$) samples which
consist of 56 blue and 56 red clusters are shown in the right panels.
While the blue clusters show a constant velocity dispersion of
about $200\,\textrm{km\,s}^{-1}$, the velocity dispersion of the red
clusters drops from a value of $\sim\!240\,\textrm{km\,s}^{-1}$ to
$\sim\!170\,\textrm{km\,s}^{-1}$ at a radius of about
2\arcmin-3\arcmin. From this radius on, out to at least $5\farcm\,5$,
the velocity dispersion apparently remains at the constant, low
value.
Considering these results for blue and red clusters, one can attribute
the drop seen for the whole error selected sample to the red
clusters. 
However, the number of red clusters is rather high in the innermost
radial bins, giving the large dispersion at small radii a high degree
of confidence. On the other hand, the lower velocity dispersion at
larger distances may be uncertain for the individual bins, but that
the overall velocity dispersion is lower than for the innermost bin
can be stated with high confidence as well. This is a new feature
which has not been seen before in any of the few kinematic GC samples
of other galaxies. We offer a simple explanation in the context of the
spherical Jeans equation in the next section.
\begin{table}
\begin{center}
\begin{tabular}{c r@{$\pm$}l r@{$\pm$}lr@{$\pm$}l}
\hline
\hline
\multicolumn{1}{c}{$R$} & 
\multicolumn{2}{c}{$\sigma$\,(all)}& 
\multicolumn{2}{c}{$\sigma$\,(blue)}&  
\multicolumn{2}{c}{ $\sigma$\,(red)}
\\
\multicolumn{1}{c}{(arcmin)} & 
\multicolumn{2}{c}{$(\textrm{km\,s}^{-1})$}&  
\multicolumn{2}{c}{$(\textrm{km\,s}^{-1})$}&  
\multicolumn{2}{c}{$(\textrm{km\,s}^{-1})$}    
\\ 
\hline
\multicolumn{7}{c}{no error selection}\\
\hline
1.7 & $228$& $23$ & $218$& $30$ &  $241$& $39$ \\
3.2 & $157$& $22$ & $160$& $33$ &  $155$& $32$ \\
4.7 & $204$& $24$ & $205$& $33$ &  $207$& $37$ \\
6.2 & $206$& $25$ & $196$& $30$ &  $190$& $39$ \\
6.2 & $(169$& $21)$ &$(126$&$21)$&  \multicolumn{2}{c}{\,} \\
7.7 & $207$& $43$ & $220$& $63$ &  $196$& $65$ \\
9.2 & $105$& $41$ &  \multicolumn{2}{c}{\,} & \multicolumn{2}{c}{\,} \\
\hline
\multicolumn{7}{c}{error-selected samples $(\Delta \varv < 35 \textrm{km\,s}^{-1})$}\\
\hline
1.7 & $221$& $25$ & $198$& $32$ &  $241$& $38$ \\
3.2 & $185$& $37$ & $207$& $64$ &  $172$& $46$ \\
4.7 & $175$& $24$ & $181$& $34$ &  $168$& $33$ \\
6.2 & $226$& $34$ & $227$& $46$ &  $194$& $42$ \\
6.2 & $(178$& $27)$ & $(121$&$27)$& \multicolumn{2}{c}{\,}     \\
7.7 & $162$& $260$ & \multicolumn{2}{c}{\,} &\multicolumn{2}{c}{\,}        \\
9.2 & $105$& $41$  & \multicolumn{2}{c}{\,}&\multicolumn{2}{c}{\,}         \\
\hline
\hline
\end{tabular}
\end{center}
\caption{Projected velocity dispersion as function of projected
galactocentric radius. The upper section shows the values for all, the
blue and the red clusters. The dispersions for the corresponding
error-selected samples are listed in the lower section. The data are
shown in Fig.~\ref{dispall}. The values in brackets are those obtained
when omitting the two (blue) GCs with velocities above
$1400\,\textrm{km\,s}^{-1}$.}
\label{disptab}
\end{table}
%
\section{The Mass Profile}
\label{chap:dynmass}
The line-of-sight velocity dispersion $\sigma_{los}(R)$ of the globular
cluster system and its radial dependence is used to estimate the mass
profile of NGC\,4636. Given that the number of spectroscopically
observed GCs is still far too low for an axisymmetric treatment, we
will assume spherical symmetry. This is justified since the deviation
from sphe\-ricity is modest in the inner parts of the galaxy (Dirsch et
al.~\cite{boris}) where most of our clusters are found.
A further argument is the high ${M}/{L}_{\mathrm{B}}$ found by Kronawitter et
al.~(\cite{2000A&AS..144...53K}) (see \ref{sect:luminous}) Given that
a flattening in the line of sight leads to an underestimation of $M/L$ 
(Magorrian \& Ballantynes \cite{magorrian01}), it is hard to imagine
that such a high value is still underestimated.
\subsection{The Mass Profile from the Jeans Equation}
\label{sect:jeans}
The collisionless Boltzmann equation is used to derive the
spherical, non-rotating {Jeans equation} 
(see Binney~\& Tremaine \cite{BT} for a full treatment).
\begin{equation}
\frac{\mathrm{d}\left(\ell(r)\, \sigma_{r}(r)\right)}{\mathrm{d}r} +2\, \frac{\beta(r)}{r}\,\ell(r)\,\sigma_{r}(r)= - \ell(r) \,\frac{G\cdot M(r)}{r^{2}}
\label{eq:jeans}
\end{equation}
\begin{displaymath}
\textrm{with}\qquad \beta \equiv 1 - \frac{{\sigma_{\theta}}^2}{{\sigma_{r}^2}}\;.
\end{displaymath} 
Here, $r$ is the radial distance from the center and $\ell$ is the
spatial (i.e.~three--dimensional) density of the GCs; ${\sigma_r}$ and
${\sigma_\theta}$ are the radial and azimuthal velocity dispersions,
respectively. $\beta$ is the anisotropy parameter, and $G$ is the
constant of gravitation.  To derive the enclosed mass $M(r)$, one
needs to know $\beta$. In other words: the potential and the
anisotropy are degenerate in the spherical approximation.
Information about the anisotropy must be inferred from higher moments
of the velocity distribution, for which one needs much larger samples
than ours. Merritt~\& Tremblay (\cite{1993AJ....106.2229M}) used M\,87
as an example and estimated that several hundred, perhaps thousand
velocities would be required to detect a velocity anisotropy.
\subsection{The Jeans--Analysis: Principle }

For our analysis we use the expressions given by e.g.~Mamon \&
{\L}okas (\cite{lundm}) (see also van der Marel (\cite{marel94})).
Given a mass distribution $M(r)$, a three--dimensional mass or number
density of a tracer population $\ell(r)$, and a \emph{constant}
anisotropy parameter $\beta$, the solution to the Jeans equation (Eq.~\ref{eq:jeans})
reads:
\begin{equation}
\ell(r)\, {\sigma_{r}^{2}(r)} =  \mathrm{G} \int_r^\infty \ell(s) \, M(s) \frac{1}{s^{2}} \left( \frac{s}{r} \right)^{2 \beta} \mathrm{d} s
\label{eq:sigl}
\end{equation}
This expression is then projected using the following integral:
\begin{equation}
\sigma_{\mathrm{los}}^{2}= \frac{2}{I(R)} \left[ \int_{R}^{\infty}  \frac{\ell \sigma_{r}^{2}\, r\, \mathrm{d}r}{\sqrt{r^{2}-R^{2}}}
- R^{2}\int_{R}^{\infty}  \frac{\beta \ell \sigma_{r}^{2}\,  \mathrm{d}r}{r \sqrt{r^{2}-R^{2}}} \right]
\label{eq:siglos}
\end{equation}
where $I(R)$ is the projected number density of the tracer population.
\subsection{Luminous Matter}
\label{sect:luminous}
To assess the mass
distribution of the luminous (stellar) component, we proceed as
follows: The photometric study by Dirsch et al.~(\cite{boris})
provides a surface brightness profile in Kron-Cousins R. It reads:
\begin{equation}
 \mu(R) = -2.5 \log \left( a_1 \left(1+\frac{R}{R_1} \right)^{-\alpha_1}
+ a_2 \left(1+\frac{R}{R_2} \right)^{-\alpha_2}
 \right)\;,
\end{equation}
with $a_1= 3.3 \times 10^{-7}$, $R_1=0.11\arcmin$, $\alpha_1 = 2.2$,
$a_2=5.5 \times 10^{-9}$, $R_2 =8.5\arcmin$, and $\alpha_2= 7.5$.  To
obtain the luminosity density $[L_{\sun}/\mathrm{pc}^3]$, the
intensity profile $I(R)=10^{-0.4\,\mu(R)}$ has to be deprojected.
This was done using the Abel deprojection integral
(Eq.~\ref{eq:depro}). For the cutoff--radius $r_\mathrm{t}$, we chose
a value of 2000\arcsec, roughly twice the distance of the last
photometric datapoint.
\begin{equation}
\varepsilon(r) = - \frac{1}{\pi} \left[ 
\int_r^{\,r_\mathrm{t}} \frac{d I(R)}{d R} \frac{d R}{\sqrt{R_{\rule{0ex}{1ex}}^{2} - r^{2}}} - 
\frac{I(r_{t})}{\sqrt{r_{t}^{2}-r^2}}
\right]
\label{eq:depro}
\end{equation}
Further assumptions are spherical symmetry, and a distance of
15\,Mpc. Moreover, we have to take into account that we use an
R-surface brightness profile: in order to translate it into solar
luminosities, we need the absolute magnitude of the Sun in this
filter. We use a value of $V-\!R = 0.370$ (Bessell et
al.~\cite{1998A&A...333..231B}) for the Sun,
i.e.~${M_{\mathrm{R},\sun}} = 4.46$.  The deprojected luminosity
density was then (numerically) integrated to yield a cumulative
luminosity (in units of $\mathrm{L}_{\odot}$). The latter was fit by a
sixth--order polynomial $\sum m_i\cdot R^{\,i}$ (R in parsec) with the
coefficients: $m_{0}= 6.38\times 10^{ 8}$, $m_{1}= 3.22\times 10^{ 6}$,
$m_{2}= -70.0$, $m_{3}= 5.62\times 10^{-4}$, $m_{4}=
3.54\times 10^{-9}$, $m_{5}= -8.92\times 10^{-14}$ and  $m_{6}= 4.22\times
10^{-19}$. 
 The polynomial is a good representation in the range 
$2<R<70\,\textrm{kpc}$. \\
To convert the cumulative 
luminosity into a mass profile, the stellar mass-to-light ratio is
required, i.e.~$M/L_{\mathrm{R}}$.  Kronawitter et
al.~(\cite{2000A&AS..144...53K}) quote a value of $M/L_{\mathrm{B}} =
9$ for a distance of 20.5\,Mpc. Because $M/L$ is inversely
proportional to the distance, one has to apply a factor of 1.36 to get
$M/L_{\mathrm{B}} = 12.2$ at a distance of 15\,Mpc. This value is
suspiciously high. Although it formally fits old, metal-rich
populations with steep mass functions, Bell et al.~(\cite{bell2003})
quote $M/L_{\mathrm B} =10$ as an upper limit found for their large
2MASS-sample. According to their relation of (B-R) versus $M/L_B$,
NGC\,4636 would have an $M/L_{\mathrm{B}}$ of approximately 6. Further
uncertainties are the distance (Dirsch et al.~\cite{boris}) and the
tangential anisotropy (Kronawitter et al.~\cite{2000A&AS..144...53K})
of NGC\,4636.
Obviously, there is some freedom in adopting
the mass-to-light-ratio of the stellar population.
First we assume $M/L_{\mathrm{B}} = 12.2$, according to Kronawitter et
al., but we shall also consider other values.  In order to transform
this mass-to-light ratio into an $M/L_{\mathrm{R}}$ value, one has to
apply the factor of $10^{-0.4 ((B-R)_{4636} - (B-R)_{\sun})}$. With
$(B-R)_{4636}= 1.6$ and $(B-R)_{\sun}= 0.97$, a factor of 0.56
results, hence $M/L_{\mathrm{R}} = 6.8$.  Further, we make the
assumption that $M/L_{\mathrm{R}}$ of the stellar light is constant
for the entire galaxy. This is justified since the color gradient of
NGC\,4636 is relatively shallow ($\Delta(C-T)\simeq -0.1$ in the range
$0\farcm 5 - 4\arcmin$), and the stellar populations seem to be well
mixed (Dirsch et al.~\cite{boris}).  

For our modeling (cf.~Sect.~\ref{sect:jeansana}), we assume that the
stellar mass is contained within 70\,kpc, yielding $\rm
M_{\mathrm{stars}} = 4.2 \times 10^{11}$ {and} $2.6 \times 10^{11}
\textrm{M}_{\sun}$ for $M/L_{\mathrm R}=6.8$ and $ 4.3$,
respectively. The lower value shall illustrate the dependence on
$M/L$. We adopt
$M/L_{\mathrm V}=5.5$ from Loewenstein~\& Mushotzky
(\cite{loewenstein03}) which results in $M/L_{\mathrm R}=4.3$.
\subsection{Blue GCs as  Tracer Population}
Since the projected number density  $N(R)$  of the blue globular clusters 
follows a double power--law, the de--projection is somewhat tricky. We
proceeded as follows: following
Zhao (\cite{zhao97}) we adopt a double power--law as
\begin{equation}
N(R) = n_{0} \left( \frac{R}{B} \right)^{-\gamma_1} 
\left[ 1 + \left(\frac{R}{B}\right)^{{1}/{\alpha_{1}}}  \right]^{
- (\beta_1 -\gamma_1)\alpha_1 
}\!.
\label{eq:doublepl}
\end{equation}
Capitals mean values in projection. $\gamma_1, \beta_1, \alpha_1$ are
the slope of the inner power--law, the slope of the outer power--law, and the
width of the transition region, respectively. B is a scale radius. This function was
fitted
to the number--counts, resulting in $ \mathrm n_0 = 8.02\,\textrm{GCs} \times \textrm{arcmin}^{-2}$, $B = 8.86\arcmin$,
 $\gamma_1 = 0.33$, $\beta_1 = 7.19$, $\alpha_1 = 0.54$. 

This function was then deprojected using
Eq.\,\ref{eq:depro} and in turn fitted
with the same function, resulting in $\mathrm n_0 = 0.38\,\textrm{GCs}\times \textrm{arcmin}^{-3}$, $b = 8.86\arcmin$, 
$\gamma_1 = 1.0$,  $ \beta_1 = 7.71$, $\alpha_1 = 0.4$.

 As a check, this 3D-profile was
re--projected, using the Abel projection operator

\begin{equation} 
I(R) = 2 \,\int_{R}^{r_{t}} \frac{\varepsilon(r) \,r\, \mathrm{d}r}{\sqrt{r^{2}-R^{2}}}
\end{equation}
and found to be in good agreement with
our 2D data (see Fig.~\ref{fig:profblue}).

\begin{figure}
{\resizebox{0.45\textwidth}{!}{\includegraphics[]{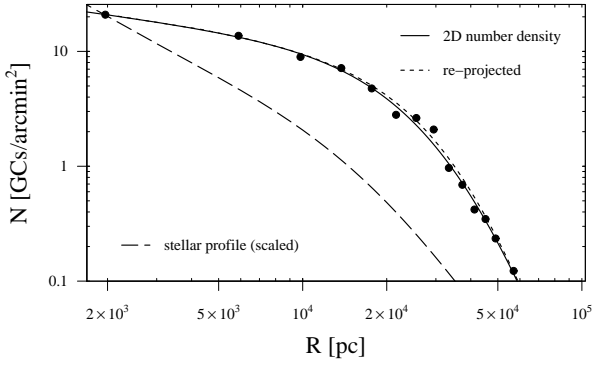}}}
\caption{Projected radial number density of blue GCs. The data points
are from Dirsch et al.~(\cite{boris}), the solid line is the a double
power--law (cf.~Eq.~\ref{eq:doublepl}) fit to the data and the dashed
line is the re-projected 3D number--density. It deviates only
marginally from the 2D--fit. For comparison we also show the projected 
light profile according to Dirsch et al. (\cite{boris}).}.
\label{fig:profblue}
\end{figure}

For simplicity, the anisotropy parameter $\beta$ is assumed to be constant
with radius. With our limited sample we cannot put any constraints on possible
 radial variations of $\beta$.

As a tracer population we choose the error-selected blue clusters. The red
clusters show a more complex behavior. They seem to rotate, their projected
number density is best described by three power-laws, and the projected
velocity dispersion also shows more complex features.  Therefore, we postpone
any inclusion of the red clusters in the dynamical analysis until more data
becomes available.

But also the blue clusters pose problems. As can be seen in Fig.11 and Fig.9,
the velocity dispersion value of the outermost bin is strongly affected by two
outliers. Including those objects one obtains $227 \pm 46\, \textrm{km\,s}^{-1}$
instead of $121 \pm 27\,\textrm{km\,s}^{-1}$. The high value would mean an
almost constant dispersion out to 7 arcmin. As seen later on this would be
inconsistent with a sharp drop of the cluster system at this radius. This
unsatisfactory situation can as well only be remedied with future data. For
now, we calculate our models for both values. 
This argument cannot be transferred easily to
the red clusters. If they rotate, a constant velocity dispersion is plausible
since we average azimuthally.

\subsection{Dark matter}
  
For the dark component we adopt a
density profile of the form:
\begin{equation}
\varrho_{\mathrm{dark}}= \varrho_{\mathrm{dark},0} \,\Big(
\frac{r}{r_{\mathrm{dark}}}\Big)^{-1} \,\Big(1+
\frac{r}{r_{\mathrm{dark}}}\Big)^{-2}\;.
\end{equation}
i.e.~an NFW profile (Navarro et al.~\cite{navarro1997}, Bullock et al.~\cite{bullock2001}).
For this density profile, the cumulative mass $M(r)$ reads
\begin{equation}
  M_{\mathrm{dark}}= 4\pi\cdot\varrho_{\mathrm{dark},0}\cdot
  r_{\mathrm{dark}}^3\cdot\Bigg(
  \log\Big(1+\frac{r}{r_{\mathrm{dark}}}\Big)
  -\frac{\frac{r}{r_{\mathrm{dark}}}}
  {1+\frac{r}{r_{\mathrm{dark}}}}
  \Bigg)\;.
\end{equation}
The task is now to find values for the characteristic density
$\varrho_{\mathrm{dark},0}$ and the characteristic radius
$r_{\mathrm{dark}}$ resulting in a dark component
which, when added to the luminous matter can
reproduce the observed projected velocity dispersion in the radial range of the
globular clusters.
\subsection{Results}
\label{sect:jeansana}
\begin{table*}
\caption{Parameters of the dark matter halos. The first columns labels
the models. The mass--to--light ratio $\textrm{M/L}_{\mathrm{R}}$ and
the dark matter fraction are calculated for two radii
($7.5\,\textrm{kpc} \simeq \rm R_{\mathrm{eff}}$\, and $30\,\rm kpc$
is the radius within which the GCs used in the analysis are found).
From the NFW parameters of our model halos, we calculate the virial
radius, mass and the {{concentration parameter}} 
${{(c_{\rm{vir}}=
\mathrm{R}_{\mathrm{vir}}/r_{\mathrm{dark}})}}$ using the definitions
of Bullock et al.~(\cite{bullock2001}). Then we give the total mass
within 30 kpc, mass-to-light ratios (R-band) and the dark matter
fraction within 30 kpc and 7.5 kpc (effective radius).
Note that the virial
quantities were calculated for the dark halos  without adding the stellar component. Thus, the
virial masses presented below are lower limits.  } \centering
\begin{tabular}{crrlclllcllll}
\hline \hline \# & $\beta$ & $r_{\mathrm{dark}}$ &
 $\varrho_{\mathrm{dark},0}$ & r.m.s & $\mathrm{R}_{\mathrm{vir}}$ &
 $\mathrm{M}_{\mathrm{vir}}$ & $c_{\mathrm{vir}}$ &
 $\mathrm{M}_{\mathrm{< 30 kpc}}$ &
 $\textrm{M/L}_{\mathrm{R}}$ & $\frac{\rm M_{\mathrm{dark}}}{\rm M_{\mathrm{total}}}$& $\textrm{M/L}_{\mathrm{R}}$ &
 $\frac{\rm M_{\mathrm{dark}}}{\rm M_{\mathrm{total}}}$\\ & & [kpc] & {\small{$[\textrm{M}_{\odot}\,
 \mathrm{pc}^{-3}]$}} & $[\rm km\,s^{-1}]$ & [kpc] & [$10^{12}\,
 \textrm{M}_{\odot}$] & $\quad$ & $ [10^{11}\textrm{M}_{\odot}]$ & \multicolumn{2}{c}{$(r=30\,\rm{kpc})$}& \multicolumn{2}{c}{ $(r=7.5\,\rm{kpc})$ } \\ \hline
&\multicolumn{12}{l}{$\mathrm{M/L}_{\mathrm{R}}= 6.8$, $\sigma_{\mathrm{los}}(27\,\rm kpc) = 121\rm{km\,s}^{-1}$ (cf.~Fig.~\ref{mod68}, left panels)\rule[-1.ex]{0ex}{4.0ex}} \\ \hline
1& $- 2.0$ & 5 & 0.12 & 23 & 207 & $ 0.52 $ & 41.4 & 5.5  & 10.9 & 0.37 & 9.6 & 0.29\\  
2&$- 0.5$ & 10 & 0.033 & 20 & 255 & $ 0.96 $ & 25.5 &5.5 & 12.0 & 0.43 & 9.4 & 0.27 \\  
3&$ 0$ & 40 & 0.003 & 21 & 395 & $3.6 $ & 9.9 & 6.6   & 13.1 & 0.48 & 8.4 & 0.19 \\  
4&$0.5$ & 120 & 0.00075 & 26  & 655 & $16.6 $ & 5.5 &7.2  & 14.3 & 0.52 & 8.2 &0.17 \\  
5&$1.0$ & 240 & 0.00050 &  46 & 1100  & 77.9 & 4.57 &9.2  & 18.3 & 0.63 & 8.7 & 0.22 \\ \hline
&\multicolumn{12}{l}{$\mathrm{M/L}_{\mathrm{R}}= 4.3$,   ${\sigma_{\mathrm{los}}(27\,\rm kpc) = 121\rm{km\,s}^{-1}}$ (cf.~Fig.~\ref{mod43}, left panels)\rule[-1.ex]{0ex}{4.0ex}} \\ \hline
6&$- 2.0$ & 5 & 0.195 & 22 & 248 & $ 0.90 $ & 49.6 &5.5 & 10.9 & 0.61 &8.9& 0.52\\  
7&$- 0.5$ & 5 & 0.230 & 20 & 264 & $ 1.01 $ & 52.8 &6.1 & 12.1&0.64 &9.7&0.56\\  
8&$0$     & 20 & 0.0135 & 20 & 358 & 2.7 & 17.9 &5.3 & 10.6 & 0.59 &5.9& 0.27\\
9&$0.5$     & 120 & 0.00100 & 25 & 743 & 24.1 & 6.2 &7.2 &14.3& 0.70&6.2& 0.30\\
10&$1.0$      & 120 & 0.00145 & 45 & 870 & 39.0 & 7.3 &9.4 &18.7& 0.77&7.0&0.37\\
\hline \hline
&\multicolumn{12}{l}{$\mathrm{M/L}_{\mathrm{R}}= 6.8$,  $\sigma_{\mathrm{los}}(27\,\rm kpc) = 227\rm{km\,s}^{-1}$   (cf.~Fig.~\ref{mod68}, right panels) \rule[-1.ex]{0ex}{4.0ex}} \\ \hline
11&$- 2.0$ & 120 & 0.0012 & 18 & 803 & $30.5  $ & 6.7 &9.4 & 18.7&0.64&  9.0& 0.25\\  
12&$- 0.5$ & 80 & 0.0019 & 25 & 650 & $ 16.0 $ & 8.1 &9.0  & 26.7&0.75&   10.5& 0.35 \\  
13&$ 0.0$ & 120 & 0.0011 & 32 & 773 & $ 27.2 $ & 6.4 & 9.0  & 17.7& 0.62& 8.8 & 0.23\\  
14&$ 0.5$ & 120 & 0.0012 & 47  & 803 & $ 30.5 $ & 6.7 &9.5  &18.7& 0.64& 9.0 & 0.25\\  
\hline
&\multicolumn{12}{l}{$\mathrm{M/L}_{\mathrm{R}}= 4.3$,  $\sigma_{\mathrm{los}}(27\,\rm kpc) = 227\rm{km\,s}^{-1}$  (cf.~Fig.~\ref{mod43}, right panels) \rule[-1.ex]{0ex}{4.0ex}} \\ \hline
15&$ - 2.0 $ & 80 & 0.0023 & 16 & 703 & $ 20.4 $ & 8.8 &9.0 &18.0&0.76&7.1&0.39\\  
16&$- 0.5$ & 120 & 0.0011 & 24 & 774 & $ 27.3 $ & 6.5 &7.7  &15.2&0.72&6.3& 0.32\\  
17&$ 0.0$ & 120 & 0.00135 & 27 & 845 & $ 35.4 $ & 7.0 &9.0  &17.7&0.76 &6.8&0.37\\  
18&$ 0.5$ & 120 & 0.0014 & 42 & 858 & $ 37.0 $ & 7.15 &9.2  &18.2&0.76&6.9& 0.38\\  
\hline\hline
\end{tabular}
\label{tab:halos}
\end{table*}
Having chosen a value for the stellar M/L and $\beta$, we calculate
according to Eq.~\ref{eq:sigl} the product of radial velocity
dispersion squared and the 3--dimensional number density. This we
project according to Eq.~\ref{eq:siglos}. For each pair of M/L and
$\beta$, we vary the dark matter parameters ($r_{\mathrm{dark}},
\varrho_{\mathrm{dark,0}}$), until the projected model dispersion
shows minimal residuals with regard to the observed dispersion.  For
simplicity, the four data points were given equal weights and the residuals
were calculated using
\begin{equation}
\rm r.m.s = \sqrt{\frac{\sum (\sigma_{\mathrm{model}} - \sigma_{\mathrm{los}})^{2} }{n}}\,. 
\end{equation}
We defined a set of radii $(r_{\mathrm{dark}}
\, \epsilon\,  \{ 5,10,40,80,120,240\}\, \rm{kpc})$  for which we then
adjusted the density.  Table\,\ref{tab:halos} lists the model
parameters together with derived quantities such as the virial mass and
radius, the $\mathrm M/L_R$-values, and the dark matter fractions within
30 kpc and 7.5 kpc (the effective radius).
We
determined the virial radius according to the Bullock et
al.~definition (i.e.~at $R_{\mathrm{vir}}$, the mean density of the
halo is 337 times the mean mass density of the universe,
$\rho_{0}\cdot\Omega_{\mathrm{m}}$).
Note that the virial quantities were calculated for the dark
component alone, with the aim of comparing their properties to those of
simulated N--body halos (cf.~Sect.~\ref{sect:bullock}). Thus, having
omitted the stellar component, the virial masses are
lower limits.

Our models are displayed in Figs.~\ref{mod68} and \ref{mod43}, which
show our analysis for the two different stellar $\mathrm M/L_R$-values of
6.8 and 4.3, respectively.  Table\,\ref{tab:halos} lists the halo
parameters.  The left and right panels refer to the difference in the
last data bin as seen in the topmost panels which show the comparison
between our models and the data. The observed velocity dispersion
together with their statistical uncertainties are indicated. We
emphasize the importance of the uncertain outermost bin which is
clearly dominated by systematics as mentioned above.  The various
anisotropic models are distinguished by different line styles which
can be identified in the second panel which shows the associated dark
matter circular velocities.  The degeneracy of potential and
anisotropy is nicely visible. Only the models without dark matter,
which are shown separately in Fig.~\ref{fig:nodm} seem to be excluded.

\begin{figure*}
\centering
    {\includegraphics[height=\textwidth,angle=270]{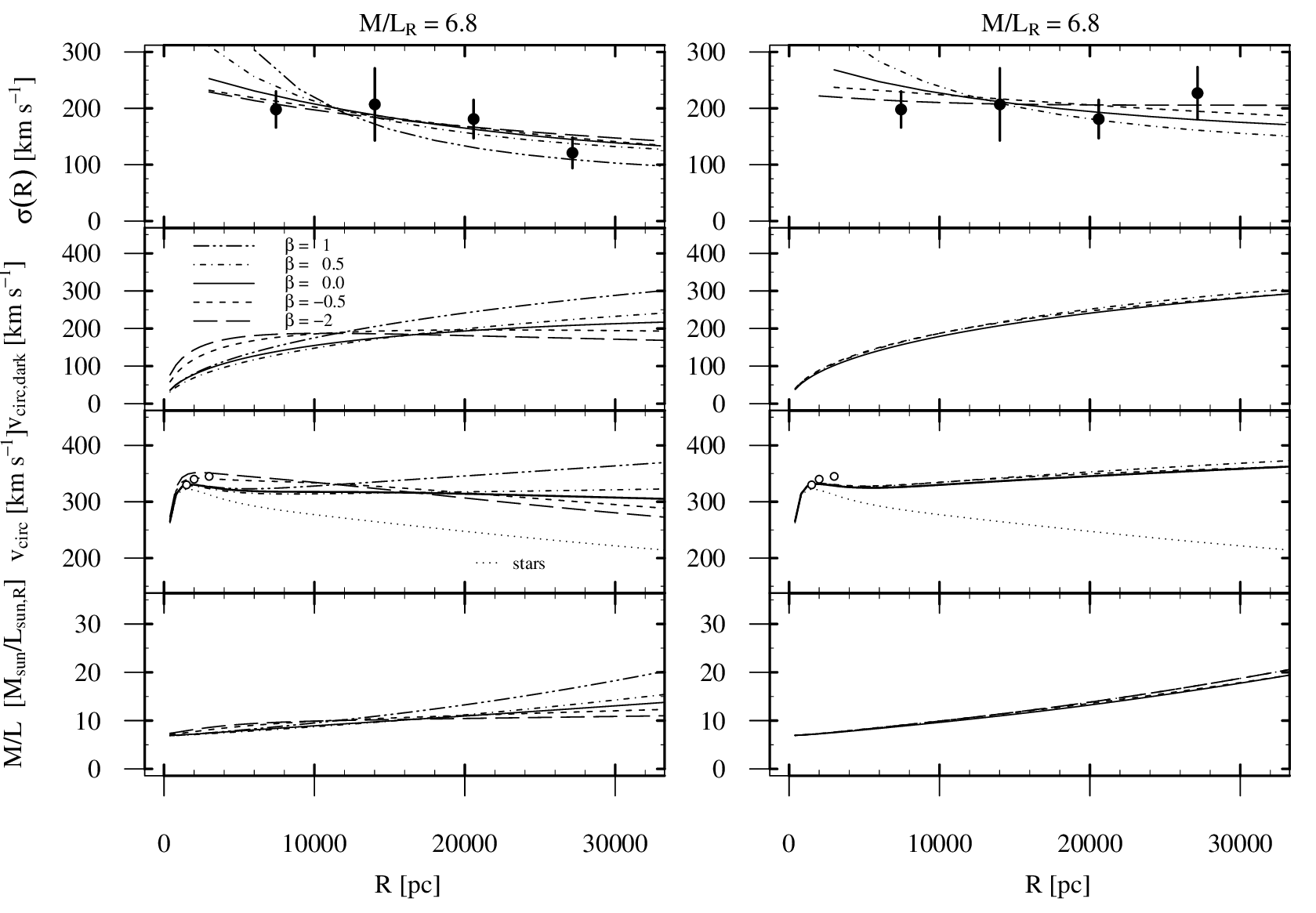}}
    \caption{Mass models for a stellar M/L--ratio of 6.8. The left and
    right panels are distinguished by the value of the last bin. The
    topmost panels show the measured velocity dispersions
    (cf.~Fig.~\ref{dispall} and Table\,\ref{disptab}) together with
    the modelled line--of--sight velocity dispersions. The second
    panels show the circular velocity curves of the dark matter
    components. Below that, the circular velocity curves for the sum
    of luminous and dark matter are shown. The circular velocity of
    the stars alone is shown as dotted line. Circles show the values
    derived by Kronawitter et al.~({\cite{2000A&AS..144...53K}}). The
    mass to light--ratio as a function of radius is shown in the
    bottom panels. Different line types indicate results for different
    anisotropy parameters as labeled in the second panel on the
    left.}
    \label{mod68} 
\end{figure*}

\begin{figure*}
  \centering
      {\includegraphics[height=\textwidth,angle=270]{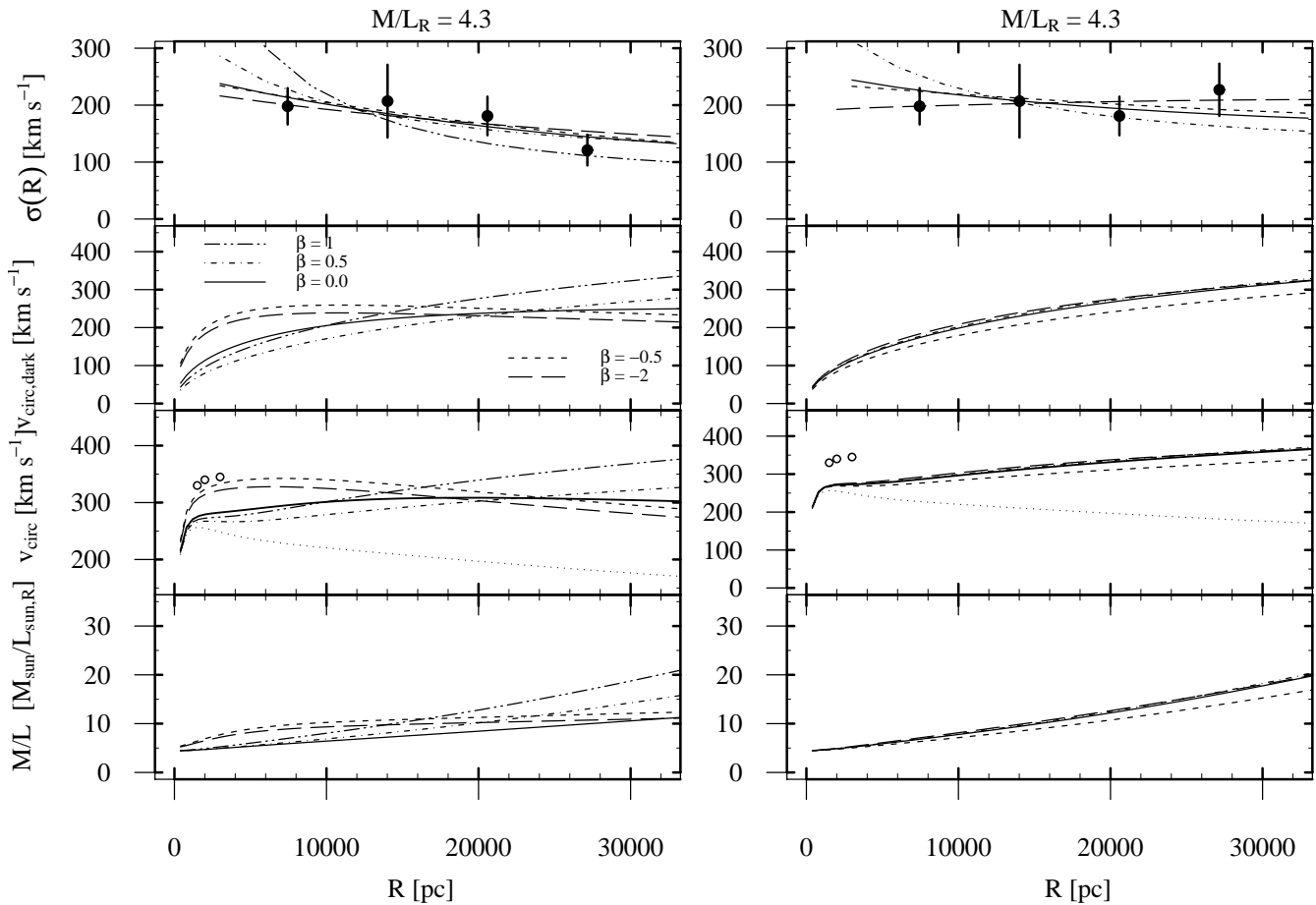}}
      \caption{Same as Fig.~\ref{mod68}, but for a lower stellar M/L--ratio of 4.3. See Table\,\ref{tab:halos} for the model parameters.}
\label{mod43}
\end{figure*}

\begin{figure}
  \centering
      {\includegraphics[width=0.49\textwidth,angle=0]{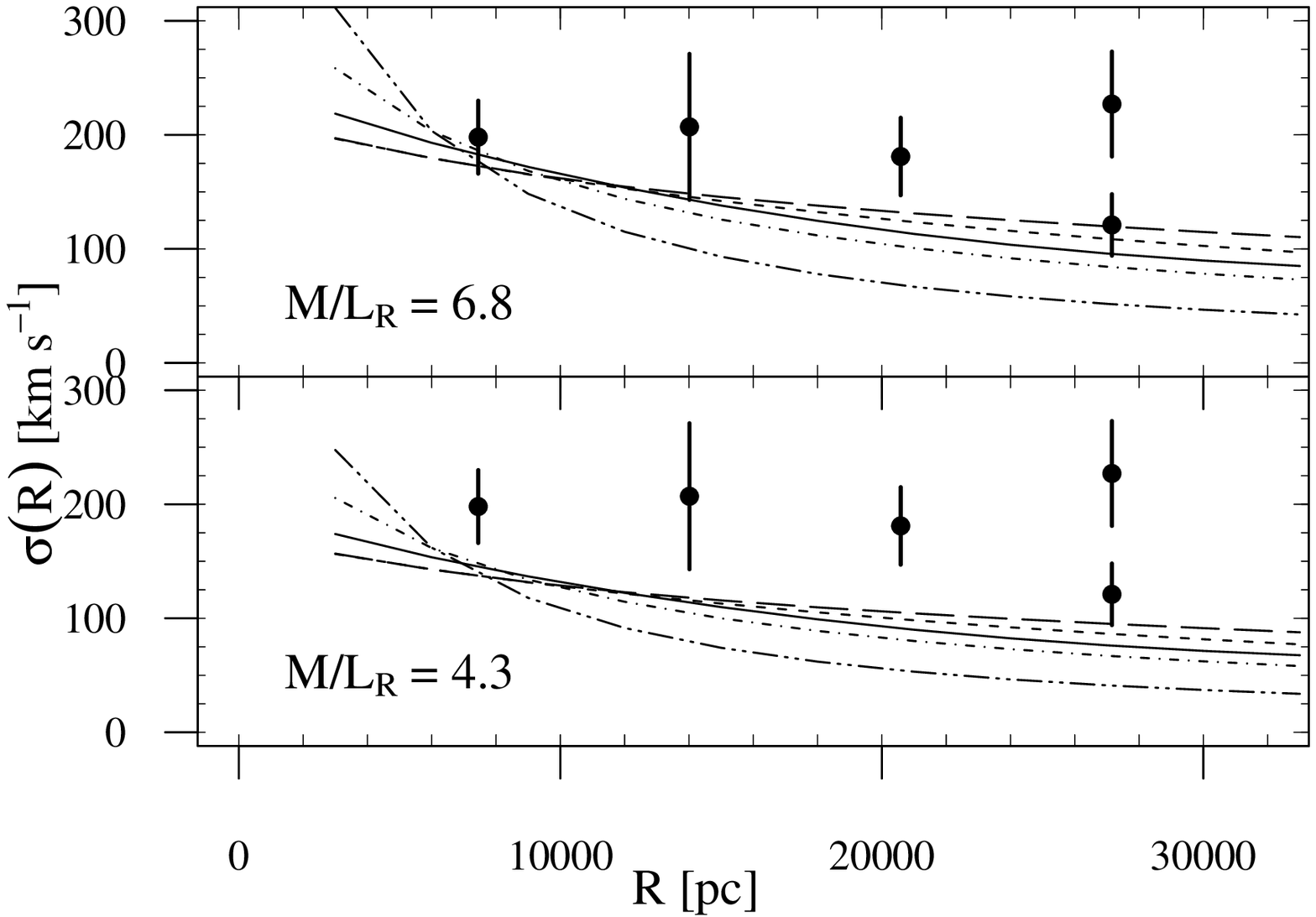}}
      \caption{Projected velocity dispersions for models \emph{without} dark
      matter. Upper panel: $M/L_{R} = 6.8$, lower panel: the same for
      a stellar mass--to--light ratio of $M/L_{R}= 4.3$. The line
      styles are the same as in Figs.~\ref{mod68} and \ref{mod43},
      i.e.~the solid lines correspond to $\beta = 0$, the long and
      short dashed lines to the tangential models $\beta =
      -\frac{1}{2}, -2$, respectively. The radial models $\beta =
      \frac{1}{2}, 1$ are shown as dot--dashed and dot--dot--dashed
      lines. Only the models with a high stellar mass--to--light ratio
      in conjunction with tangential anisotropies are marginally
      consistent with the measured line--of--sight velocity
      dispersions.}
      \label{fig:nodm}
\end{figure}
The third panels show the total circular velocities, i.e.~the sum of
dark and luminous matter. The stellar component is indicated with a
dotted line. The circular velocity values derived by Kronawitter et
al.~are shown as circles.  Since our stellar mass--to--light ratio of
6.8 was adopted from the Kronawitter et al.~analysis, it is not
surprising that they do not fit well in the case of M/L=4.3.

The bottom panels show the radial dependence of the
corresponding $\mathrm M/L_R$-values (cf.~Table\,\ref{tab:halos} for a
listing of mass--to--light ratios at $R= 30$ and $7.5$\,kpc).  

Of course, since the stellar M/L-value enters as a free
parameter, the dark matter fraction would rise with declining M/L.
However, $\mathrm M/L_R = 4.3$ is already low for an old, metal-rich
stellar population. Further constraints on the M/L ratio have to come
from a more precise analysis of the dynamics of the stellar body.

In all four model families, the $\beta=1$ (i.e.~totally radial) models
perform worst. Besides being a poor fit to the data, these models
require very massive ( $ \rm M_{\textrm{vir}} > 10^{13} \rm M_{\sun}$)
and extended dark matter halos.  

The observed GC velocity dispersion
(calculated omitting the two extreme velocities) as shown in the left
panels of Figs.~\ref{mod68} and \ref{mod43} can be reproduced for a
range of anisotropy parameters ($ -2 < \beta < 0.5$). The 
tangential models result in less massive and more concentrated dark
halos. The isotropic model has a virial radius of about 360--400\,kpc
(depending on the stellar M/L value) which is compatible to the extent
of the X--ray emission. 

Although the viral masses of the dark halos differ by more than one 
order of magnitude, the total mass within the radial range
probed by the GC dynamics changes by only a factor of $1.5$. 

Hence, we can put  constraints on the amount of dark matter in
NGC\,4636 within 30\,kpc $(\simeq 4 \,  R_{\mathrm{eff}})$, while
the shape of the halo remains largely unknown. 

For completeness we also included in our analysis the values of
$\sigma_{\mathrm{los}} (R)$ obtained when including extreme velocities
of two GCs in the last radial bin.  The resulting almost constant
line--of--sight velocity dispersion of the GCs as shown in the right
panels of Figs.~\ref{mod68} and \ref{mod43} can only be reproduced
with a very extended dark matter component.  The best agreement in the
sense of lowest residuals is found for the $\beta = -2$ tangential
models. The radial $\beta = 0.5$ models provide rather poor fits, and
the case of completely radial orbits $\beta = 1$ (not shown) appears
even worse.

\section{Discussion}
\subsection{Comparison to Cosmological Simulations }
\label{sect:bullock}
How do our model dark matter halos compare to cosmological N--body
simulations? Is such a comparison meaningful? Particularly the less
massive halos are expected to have experienced contraction during the
galaxy formation process.  On the theoretical side, the universality
of dark matter halos is still under discussion (see Reed et
al. \cite{reed05} and references therein) so that a possible
disagreement may not provide a solid basis for deeper conclusions. We
therefore refrain from lengthy considerations and only point out basic
tendencies.\\
For comparison with simulations we choose the study of Bullock et
al.~(\cite{bullock2001}). They analyzed a sample of $\sim\!5000$
simulated halos in the range $10^{11}-10^{14} \textrm{M}_{\sun}$. They
found (like others) that the concentration parameter
$c_{\mathrm{vir}}$ decreases with growing mass, but the scatter in the
halo profiles for a given mass is still quite large. A comparison with
Fig.4 of Bullock et al. shows that for our model family shown in the
left column of Fig.~\ref{mod68} (i.e.~$M/L_{\mathrm{R}}=6.8$, omitting
the two extreme blue clusters), the $\beta = 0$ halo (model\,3) fits
best to the simulations. The halos corresponding to the models with
tangential bias have higher concentration parameters than halos of the
same mass shown by Bullock et al., but may be still consistent. The
models with radial anisotropies have small concentration parameters,
but are as well not excluded. This is true even for the completely
radial models (which give the worst fits to the observations).  The
question therefore is whether, given the large range of permitted halo
shapes indicated by the simulations, one can meaningfully compare at
all theoretical simulations with observations of our kind, not to
mention the complications if dissipative effects of galaxy formation
are taken into account.  All we can say is that our model halos
do not contradict the simulations of Bullock et al.

\subsection{Mass Profile and Comparison with X-ray Mass Profiles}
NGC 4636 has recently been a target for various X-ray studies (e.g.~
O'Sullivan et al.~\cite{osullivan05}, Ohto et al.~\cite{ohto03}, Xu et
al.~\cite{2002ApJ...579...600X}, Jones et
al.~\cite{2002ApJ...567L.115J}), emphasizing the highly disturbed
X-ray pattern. However, mass profiles derived from X-rays are until
now published only by Matsushita et al.~(\cite{1998ApJ...499L..13M})
and Loewenstein \& Mushotzky~(\cite{loewenstein03}). Figure~\ref{tom3}
shows the various mass profiles. The two solid lines are our models
with the lowest and the highest mass within 30 kpc (models\,1 and
14). The dotted line is the model with the highest dark matter content
within one effective radius (model\,7).
\begin{figure}
\centering
{\resizebox{0.45\textwidth}{!}{\includegraphics[bb=215 57 565 605,clip=,angle=270]{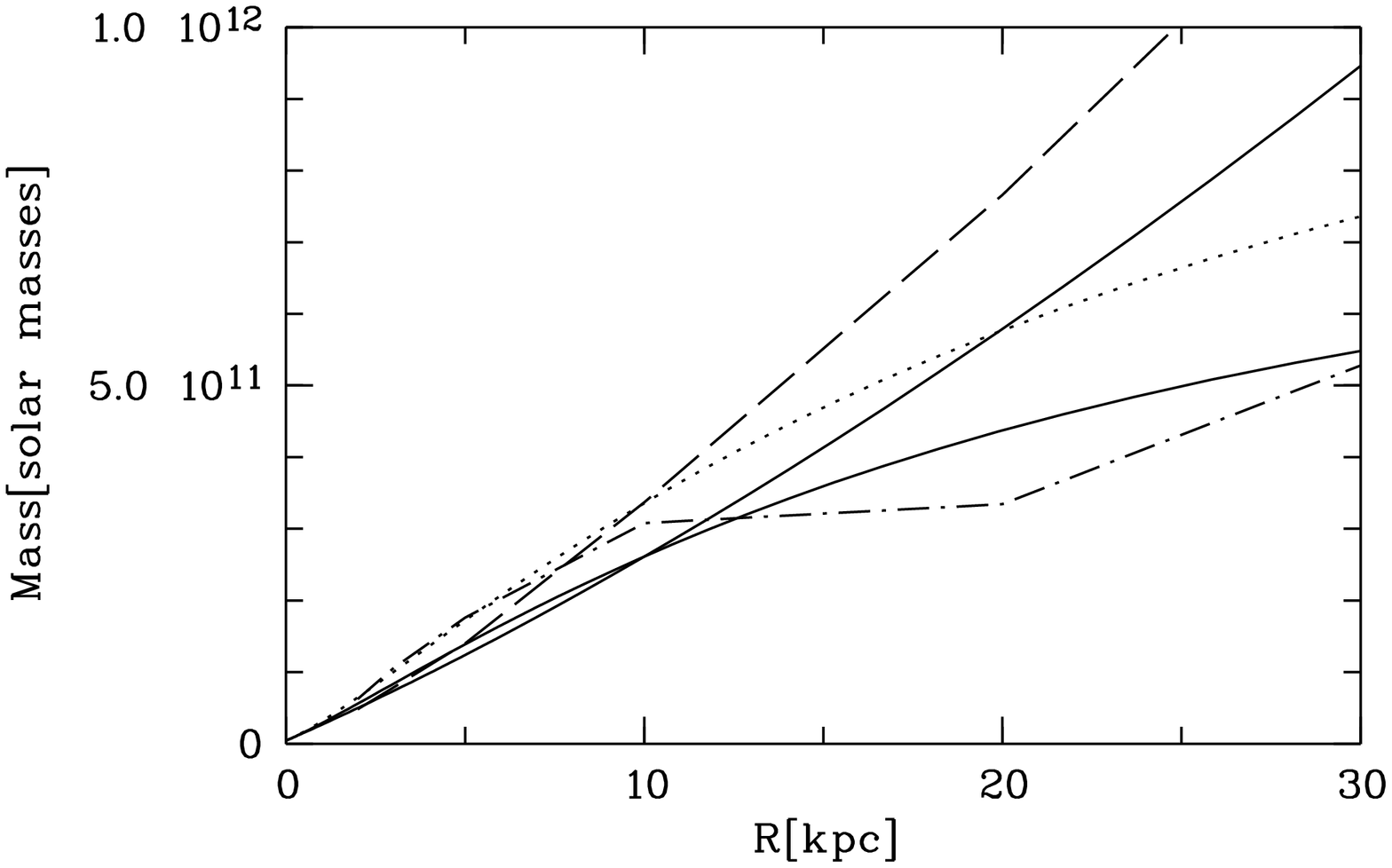}}}
\caption[Mass Profiles]{{\bf{Various mass profiles.}} This graph shows
various mass profiles of NGC 4636. The solid lines are the models with
the lowest and highest mass within 30 kpc, respectively (models\,1 and
14). The dotted line is the model with the highest dark matter content
within 30 kpc (model\,7).
The dashed dotted line approximates
one mass profile from Matshushita et al. (\cite{1998ApJ...499L..13M}),
showing the flattening at about 10\,kpc. The long--dashed line
represents the mass profile from Loewenstein \&
Mushotzky~(\cite{loewenstein03}). See text for more details.}
\label{tom3}
\end{figure}
The long--dashed line is the mass profile from Loewenstein \&
Mushotzky~(\cite{loewenstein03}). The total mass enclosed within
30\,kpc is for the latter profile $1.25\times 10^{12}\,\rm{M_{\sun}}$,
distinctly higher than our highest value. Furthermore, the mass
profile is steeper than most of our models, rising as $R^{1.2}$.
However, a very concentrated dark halo, even surpassing this X-ray
profile at small radii, is not excluded.  One can speculate that the
difference at larger radii might have to do with different
hydrodynamical situations at small and large radii.

The dashed-dotted line is the mass profile from Matsushita et
al.~\cite{1998ApJ...499L..13M} (based on ASCA data), read off from
their log-log diagram (their Fig.~3, solid line, scaled down to our
distance of 15\,Mpc) which shows most distinctly the flattening of the
mass profile at 10\,kpc radius. However, their range of mass profiles
consistent with the X-ray analyses is rather large, apparently
including the possibility that there is no such flattening.  In their
work, NGC 4636 is extremely dark matter dominated, the luminous mass
falling short by a factor of 2-3 already at small radii. Matsushita et
al.~quote M/L = 8 for the stellar component but give neither the band
nor the source of their photometry, so we cannot comment further on
that.  The agreement with the GC analysis out to 10\,kpc is reasonably
good. Outside this radius, the discrepancy is caused by the flattening
of the profile for which we see no indication.  A similar flattening
of the mass profile has been observed with ASCA in the cases of NGC
1399 (Ikebe et al.~\cite{ikebe1996}) and in Abell 1795 (Xu et
al.~\cite{2002ApJ...579...600X}).  However, it has not been seen in
the ROSAT data for NGC 1399 (Jones et al.~\cite{jones1997}). Jones et
al.~(\cite{2002ApJ...567L.115J}) interpret the high X-ray luminosity
of NGC 4636 as the transient phenomenon of an enhanced cooling rate.
Hence it is possible that the total gravitating mass of this galaxy
cannot be accurately determined from the X-ray emission.  An
interesting point is the large radial extension of the X-ray emission
found by ASCA.  If the dark matter fraction in NGC 4636 is low then
this emission might indicate a dark halo not related to NGC 4636
itself. One could speculate about a dark substructure in the general
Virgo potential, formerly hosting a galaxy group which formed NGC 4636
as the result of the merging of several galaxies.  The quite abrupt
termination of the GC system between 7\arcmin - 9\arcmin would speak
in favor of this scenario. Stars or GCs, respectively, near their escape energy
are expected to show a drop in their radial density profile. If
NGC\,4636 was embedded in a deep dark matter potential reaching
smoothly out to large radii then one would expect to find GCs at large
radii which are still bound to NGC 4636 and accordingly a sharp
boundary would not fit into the picture. 
{{However, it is interesting to note that the blue clusters apparently do not
trace the dark halo, as has been suggested for other galaxies, e.g. NGC\,1399 (Forte et al., \cite{forte05}).}}
 Deeper insight could come
from a more complete sampling of GCs or Planetary Nebulae at or beyond
a radius of 9\arcmin, which will be a difficult observational task.
\par
\subsection{Modified Newtonian Dynamics}
\begin{figure}
\centering
{\resizebox{0.45\textwidth}{!}{\includegraphics[bb=192 59 540 464,clip=,angle=270]{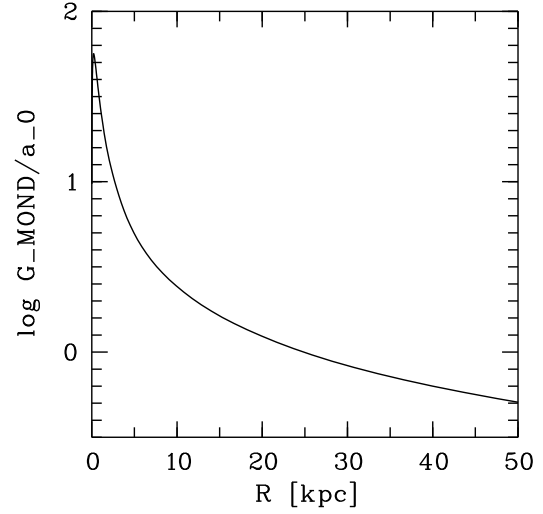}}}
\caption[]{This graph shows the ratio of the MONDian acceleration to
$a_0$ as a function of galactocentric radius. Equality is reached at
about 30 kpc which already is the limit of our investigation. This
means that the deep MOND regime cannot be reached by dynamical tracers
and that the scheme of interpolating between the MONDian and the
Newtonian regime dominates the analysis.  A stellar mass--to--light
ratio of $M/L_{\mathrm{R}}=6.8$ was assumed.}\label{newton}
\end{figure}

\begin{figure}
\centering
{\resizebox{0.48\textwidth}{!}{\includegraphics[bb= 220 65 565 585,clip=,angle=270]{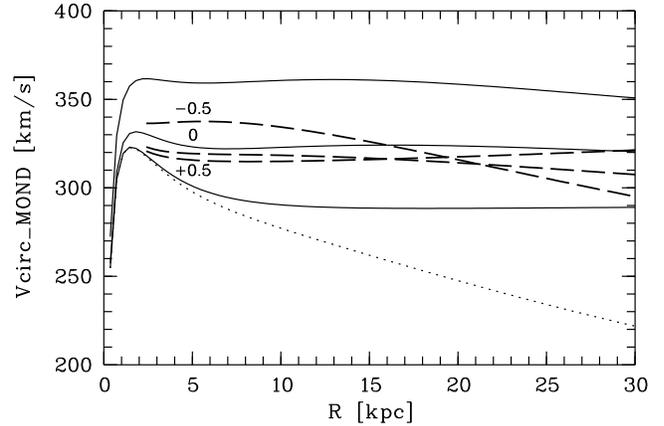}}}
\caption[]{ MOND rotation curves for different interpolation schemes
are shown for a stellar $M/L_{\mathrm{R}}$ ratio of 6.8 (solid
lines). The upper solid line is derived from Bekenstein's
interpolation, the middle one from Famaey \& Binney's, and the lower
one from the ''standard'' interpolation (see text for explanation).
The dashed lines indicate the derived circular velocities from GCs for
the given anisotropy parameters (models 3,4,5). The dotted line is the
stellar component alone with an adopted $M/L_{\mathrm{R}}=6.8$. The
Famaey \& Binney interpolation apparently works best.}
\label{MOND}
\end{figure}

The CDM paradigm is the most widely accepted view onto structure and
galaxy formation. Alternative concepts to explain the kinematics of
galaxies without resorting to dark matter are until today minority
views.  One of these alternative concepts is Modified Newtonian
Dynamics (MOND) which one finds to be discussed with increasing
intensity.  (e.g.~Milgrom \cite{milgrom83}, Sanders \& McGaugh
\cite{sanders02}).  As a phenomenological recipe, the Newtonian
acceleration ${\bf{ g_N}}$ has to be replaced by the MOND acceleration
in the case of small accelerations ${\bf g}$ through the relation
\begin{equation}
{\bf{g_N}} = \mu (g/a_0) \cdot \bf{g}\;,
\end{equation}  
with
$$ a_0 \approx 1.2 \cdot 10^{-8} \textrm{m} \cdot \textrm{s}^{-2}$$
being a universal constant and $\mu(g/a_0)$ an unspecified function
interpolating between the Newtonian and the MOND regime.  Bekenstein
\& Milgrom (\cite{beken84}) presented a Lagrangian formalism of
MOND. In case of a spherical mass distribution this formalism reduces
to the above simple formula.

 MOND proved
to have an impressive predictive power (e.g. Sanders \& McGaugh
\cite{sanders02}). Many arguments have been raised against MOND,
including its apparent arbitrariness, but until now no convincing case
has been found which would disprove MOND. Meanwhile, MOND also has a
theoretical foundation in the work of Bekenstein (\cite{bekenstein04}).
Perhaps the most serious objection is that MOND does not remove the
need for dark matter on cluster scales (Sanders \cite{sanders03},
Pointecouteau \& Silk \cite{pointe05}). However, ignoring pro's and
contra's for the moment, we want to see MOND's performance in the case
of NGC 4636.

 MOND works successfully in the cases of spiral galaxies and dwarf
galaxies. 
Giant ellipticals are more difficult test objects due to the fact that
MOND effects start to become visible  at radii where the low surface
brightness is an obstacle for reliable investigations of the galaxy's
dynamical behavior.  However, Gerhard et al.~(\cite{gerhard01}), in
their sample of 21 elliptical galaxies, found the onset of increasing
M/L-values at accelerations, which are a factor of 10 higher than in
spiral galaxies. This would remove the universal character of
$a_0$. Milgrom \& Sanders~(\cite{milgrom03}) counterargued that there
were discrepancies between Gerhard et al.'s analysis and that of
Romanowsky et al.~(\cite{romanowsky}), indicating that the last word
has not been spoken yet. Therefore, it is interesting to have a look
at the MOND rotation curve of NGC 4636, expected from the luminosity
density profile and a certain M/L-ratio.

Figure\,\ref{newton} shows the ratio of the MOND acceleration to $a_0$ in
dependence on radial distance (for the ''standard'' interpolation
function mentioned hereafter).  An $\mathrm{M/L}_{\mathrm{R}}$-ratio
of 6.8 has been adopted. The graph shows that giant elliptical
galaxies like NGC 4636 are nowhere in the deep MOND regime. Thus it is
clear that the interpolation function plays a dominant role and the
dynamics at small radii do not tell anything about MOND without
knowledge of the transition behavior at high accelerations (for which
there is no theory).  A formula which has been widely used in the case
of spiral galaxies is $ \mathrm \mu(x) = \frac{x}{\sqrt{(1+x^2)}}\;$
(Begeman et al.~\cite{begeman}).

However, Famaey \& Binney (\cite{famaey05}) find that the above
interpolation does not fit the Galactic rotation curve well and propose $
\mathrm \mu = x/(1-x)$ (but see Famaey \& Binney for more
detailed remarks on the shortcomings). This formula gives a smoother
transition to the MOND regime, i.e. MOND effects show up earlier.
These two formulas have no physical motivation other than giving satisfactory
results.

Bekenstein (\cite{bekenstein04}), in his relativistic TeVeS theory of
gravitation, which includes the MOND phenomenology in case of small
accelerations, develops for spherical symmetry the interpolation $
\mathrm \mu(x) = \frac{\sqrt{1+4x} - 1}{\sqrt{1+4x} + 1} $. This
interpolation is valid only for small and intermediate accelerations
and may not be valid for the full range of acceleration with which we
are dealing with.
 
The results are shown in Fig.~\ref{MOND}. We adopted again $M/L_R =
6.8$ (an $M/L_R$ of 4.3 would in any case not be compatible with
MOND).  The dashed lines are the circular velocities derived from GCs
with their anisotropies indicated.  The dotted line is the Newtonian
circular velocity due to the luminous component alone.  The solid
lines represent the MOND circular velocities with the uppermost one
being the interpolation of Bekenstein, the middle one Famaey \&
Binney's, and the lower one the ''standard'' interpolation.
Apparently, MOND reproduces the circular velocities derived from GC
kinematics quite well within the uncertainties. The minimal conclusion
is that NGC 4636 does not provide a counterargument for MOND. The
shape of Bekenstein's interpolation at small radii is probably
irrelevant, but also at large radii the deviation from our mass models
is quite large and becomes larger with declining M/L. The
interpolation of Famaey \& Binney works seemingly better than the
''standard'' formula, but the existent data are, however, not
sufficient to perform a critical test. A case against MOND could
perhaps emerge (it only can be falsified, not verified) if the
suspected drop of the velocity dispersion was to be confirmed in
combination with a clear tangential bias. The necessary data set must
then allow also the measurement of higher moments of the velocity
dispersion.  NGC 4636 has already a rich cluster system, is more or
less spherical, and has a relatively isolated location. 
Thus one may conclude that critical
tests are difficult to perform with elliptical galaxies.  
{{
But it is worth mentioning that so far, one of the most
interesting arguments comes from X--rays. Buote et al.~(\cite{buote02}) 
note that the ellipticities of the X--ray isophotes around NGC\,720 at larger radii are larger than 
those expected from the equipotential surfaces calculated from the visible
mass distributions. This holds for both the Newtonian and the MOND case. 
}}

\section{Conclusions}
This is the first dynamical study of the globular cluster system of
the giant elliptical galaxy NGC\,4636. Several hundred medium
resolution spectra were acquired at the VLT with FORS\,2/MXU. We
obtained velocities for 174 globular clusters in the radial range
$0\farcm\,90 < R < 15\farcm\,5 $, or $0.5\,- 9\, R_{\mathrm{e}}$ in
units of effective radius. Assuming a distance of 15\,Mpc, this
corresponds to projected galactocentric distances in the range 4 to
70\,kpc.  Due to the sharp decrease in the GC radial number density at
7\arcmin - 9\arcmin about 90 per cent of our clusters are found in the
range $1\arcmin < R < 7\arcmin 5$, i.e within 4.5 effective
radii. This implies that the dynamics beyond this radius cannot be
determined with the current data set.
Within this radius, however, we find a roughly constant velocity
dispersion for the blue clusters. 
The red clusters are found to have a distinctly different behavior: at
a radius of about 3\arcmin, the velocity dispersion drops by
$\sim\!50\,\textrm{km\,s}^{-1}\,$ to about $170\,\textrm{km\,s}^{-1}$
which then remains constant out to a radius of 7\arcmin. The cause
might be the steepening of the number density profile at
$\sim\!3\arcmin$ observed for the red clusters.
Following this line of thought, one would expect a dramatic drop in
the velocity dispersion for both red and blue clusters at
7\arcmin-9\arcmin, corresponding to the observed change in the
power-law exponent of the GC number density. The observational
difficulty will be to pick up sufficient clusters in this region of
low surface density.
In order to derive a mass profile, we perform a spherical
Jeans-analysis.  For a given stellar mass-to-light ratio, a model dark
matter halo, and an orbital anisotropy, we calculate the projected
velocity dispersion and choose those halos which minimize the
residuals to the observational data.  An important obstacle is the
uncertain outermost data bin, which permits a large variety of halos
to fit. Generally, slightly tangential models fit better than radial
ones.
Adopting the high stellar M/L value of Kronawitter et
al.~(\cite{2000A&AS..144...53K}) ($M/L_R$=6.8), we find within 30 kpc
for $\beta = -0.5$ an $M/L_R$-value of 12, corresponding to a dark
matter fraction of 27\%.  But also values as high as 26 cannot be ruled out
with certainty, depending on the treatment of the last bin.
When a lower stellar M/L value is used (Loewenstein et
al.~\cite{loewenstein03}) ($M/L_R$=4.3), the amount of dark matter
that is needed to explain the observed circular velocities increases
correspondingly. Massive dark matter halos (up to 76\% dark matter can
be required.  Neither of the models can be ruled out. An important
source of uncertainty is the sparse data at the outer radius. However,
for most of the reasonable models, we found either less dark matter
than X-ray studies with ROSAT at small radii or the halos are so
concentrated that the total mass is distinctly smaller. One should
consider that the claim for the very high dark matter fraction is
based upon X-ray luminosities that arise in a region suspected to be
out of hydrostatic equilibrium, so one must proceed with caution. The
flattening of the ASCA mass profile cannot be seen. It seems that the
dark matter problem in NGC 4636 also depends to a large part on the
uncertain M/L value of the luminous matter. A dynamical
re-investigation of the inner stellar region would therefore be
worthwhile.
Furthermore, it would be of great interest to obtain more spectra for
globular clusters in the radial range beyond $7\arcmin $ to see
whether the suspected decline in the velocity dispersion can be
confirmed. To answer the question whether the NGC\,4636 globular
cluster systems rotates, it is desirable to obtain a more uniform
angular coverage.
We also investigated whether the dynamics are consistent with Modified
Newtonian Dynamics and answer that positively. The difficulty is that
the deep MOND regime is beyond our radial coverage. Because MOND does
not predict the behavior of the circular velocity in the transition
region one has to apply physically unmotivated interpolation
formulae. The interpolation given by Famaey \& Binney
(\cite{famaey05}) seems to work best.

\begin{acknowledgements}
We thank the referee, Michael Loewenstein, for constructive criticism which 
led to an improved dynamical analysis.
Y.S. thanks the DAAD for supporting her stay in Concepci\'on and
acknowledges support from the Graduiertenkolleg ``Galaxy Groups as
Laboratories for Baryonic and Dark Matter'' and a German Science
Foundation Grant (DFG--Projekt HI-855/2). T.R.~and B.D.~acknowledge
support from the Chilean Center for Astrophysics, FONDAP
No.15010003.  
\end{acknowledgements}

\appendix
The appendix lists the heliocentric velocities of globular clusters and foreground stars. The object 
identification follows the scheme \#field.\#mask\,\#slit. The coordinates are for equinox 2000. The 
colors 
and magnitudes are from Dirsch et al.~2005. It is available in electronic form.
\section{Velocities}
\label{sect:GCs}
\label{sect:stars}

\begin{thebibliography}{}

\bibitem[1998]{1998gcs..book.....A} Ashman, K.~M.~\& Zepf, 
S.~E.\ 1998, Globular cluster systems, Cambridge astrophysics series 30, Cambridge University Press, 
Cambridge 

\bibitem[1991]{begeman} Begeman, K.~G., Broeils, A.~H. \and  Sanders, R.~H. 
\ 1991, \mnras,  249, 523

\bibitem[1984]{beken84} Bekenstein, J.~D., Milgrom, M. \ 1984 {\apj}, 286,  7

\bibitem[2004]{bekenstein04} Bekenstein, J.~D., \ 2004 {\prd}, vol. 70, Issue 8, id. 083509 

\bibitem[2003]{bell2003} 
Bell, E.~F., McIntosh, D.~H., Katz, N., \& Weinberg, M.~D.\ 2003, \apjs, 
149, 289 

\bibitem[1994]{1994MNRAS.269..785B} Bender, R., 
Saglia, R.~P., \& Gerhard, O.~E.\ 1994, \mnras, 269, 785 

\bibitem[1998]{1998A&A...333..231B} Bessell, 
M.~S., Castelli, F., \& Plez, B.\ 1998, \aap, 333, 231 

\bibitem[1985]{1985AJ.....90.1681B} 
Binggeli, B., Sandage, A., \& Tammann, G.~A.\ 1985, \aj, 90, 1681 

\bibitem[1987]{BT} Binney, J.~\&  Tremaine, S.\ 1987, Galactic Dynamics, Princeton University Press, 
Princeton

\bibitem[1978]{1978A&A....64L...3B} Bottinelli, 
L.~\& Gouguenheim, L.\ 1978, \aap, 64, L3 

\bibitem[2001]{bullock2001} Bullock J.~S., Kolatt T.~S., Sigad Y. et al. 2001,
\mnras, 321, 559

\bibitem[2002]{buote02} {{Buote}, D.~A., {Jeltema}, T.~E., {Canizares},
	C.~R. \and {Garmire}, G.~P.} \ 2002, \apj, 577, 183

\bibitem[1982]{1982AJ.....87.1165B} Burstein, D.~\& 
Heiles, C.\ 1982, \aj, 87, 1165 

\bibitem[2001]{2001ApJ...559..828C} C{\^o}t{\'e}, 
P., McLaughlin, D.~E., Hanes, D.~A. et al.\ 2001, \apj, 559, 828 

\bibitem[2003]{2003ApJ...591..850C} C{\^o}t{\'e}, P., McLaughlin, 
D.~E., Cohen, J.~G., \& Blakeslee, J.~P.\ 2003, \apj, 591, 850 

\bibitem[1991]{1991trcb.book.....D} de Vaucouleurs, 
G., de Vaucouleurs, A., Corwin, H.~G., et al.
 \ 1991, Third Reference Catalogue of Bright Galaxies, Springer, New York  

\bibitem[2003]{1399.1} Dirsch, B., Richtler, T., Geisler, D. ,et al.\ 2003, \aj, 125, 1908 

\bibitem[2004]{1399.3} Dirsch, B., Richtler, T., Geisler, D., et al.\ 2004, \aj, 127, 2114 
\bibitem[2005]{boris} Dirsch, B., Schuberth, Y., Richtler, T. \ 2005, \aap, 433, 43  

\bibitem[2001]{2001ApJ...561..751F} Fall, S.~M.~\& Zhang, 
Q.\ 2001, \apj, 561, 751 

\bibitem[2005]{famaey05} {{Famaey}, B. and {Binney}, J.}, \ 2005 \mnras, 363, 603



\bibitem[1985]{1985ApJ...293..102F} Forman, W., Jones, C., \& Tucker, W.\ 1985, \apj, 293, 102 

\bibitem[2005]{forte05} Forte, J.~C., Faifer, F. \& {Geisler}, D. \ 2005, \mnras, 357, 56

\bibitem[2001]{gerhard01} Gerhard, O., 
Kronawitter, A., Saglia, R.~P., \& Bender, R.\ 2001, \aj, 121, 1936 


\bibitem[1939]{1939PASP...51..166G} Giclas, H.~L.\ 1939, \pasp, 51, 
166 

\bibitem[1977]{1977MmRAS..84...45H} Hanes, D.~A.\ 1977, \memras, 84, 45 


\bibitem[1996]{1996AJ....112.1487H} Harris, W.~E.\ 1996, \aj, 112, 1487 

\bibitem[1981]{harris81} Harris, W.E., van den Bergh, S. 1981,
        AJ, 86, 1627

\bibitem[2003]{2003A&A...398..949I} 
Idiart, T.~P., Michard, R., \& de Freitas Pacheco, J.~A.\ 2003, \aap, 398, 
949 

\bibitem[1996]{ikebe1996} Ikebe Y., Ezawa H., Fukazawa Y. et al. 1997,
  Nature 379, 427


\bibitem[2002]{2002ApJ...567L.115J} Jones, C., Forman, W., Vikhlinin, A., et al.\ 2002, \apjl, 567, 
L115 

\bibitem[1997]{jones1997} Jones, C., Stern, C., Forman, W., et al.\ 1997, \apj, 
482, 143 

\bibitem[1994]{1994A&A...287..463K} Kissler, M., Richtler, T., Held, E.~V., et al.\ 1994, \aap,
  287, 463

\bibitem[1978]{1978AJ.....83...11K} Knapp, 
G.~R., Faber, S.~M., \& Gallagher, J.~S.\ 1978, \aj, 83, 11 

\bibitem[1983]{1983AJ.....88..260K} Krishna 
Kumar, C.~\& Thonnard, N.\ 1983, \aj, 88, 260 

\bibitem[2000]{2000A&AS..144...53K} Kronawitter, A., Saglia, R.~P.,
  Gerhard, O., \& Bender, R.\ 2000, \aaps, 144, 53

\bibitem[2001]{kundu} Kundu, A. \and Whitmore, B.~C. 
\ 2001, \aj, 121, 2950
  
\bibitem[2001]{larsen} Larsen, S.~S., Brodie, J.~P., Huchra, J.~P. et al.
\ 2001, \aj, 121, 2974


\bibitem[2003]{loewenstein03} Loewenstein, M., Mushotzky, F. \ 2003, Nuclear Physics B Proc. Suppl., 
124, 91

\bibitem[2001]{2001ApJ...555L..21L} Loewenstein, M., Mushotzky, R.~F., Angelini, L., et al.\ 2001, \apjl, 555, L21

\bibitem[2001]{magorrian01} Magorrian J., Ballantyne D. \ 2001, \mnras, 322, 702

\bibitem[2005]{lundm} Mamon, G.~A., {\L}okas, E.~L. \ 2005, \mnras, 363, 705

\bibitem[1998]{1998ApJ...499L..13M} Matsushita, K., Makishima, K., Ikebe, Y., et al.\ 1998, \apjl,
  499, L13


\bibitem[1993]{1993AJ....106.2229M} Merritt, D.~\& 
Tremblay, B.\ 1993, \aj, 106, 2229 


\bibitem[1983]{milgrom83} Milgrom, M.\ 1983, \apj, 270, 
365 

\bibitem[1986]{milgrom86} Milgrom, M.\ 1986, \apj, 302, 
617 

\bibitem[2003]{milgrom03} Milgrom, M.~\&  Sanders, R.~H.\ 2003, \apjl, 599, L25 


\bibitem[1997]{navarro1997} Navarro J.F., Frenk C.S., White S.D.M. 1997, \apj, 490, 493

\bibitem[2003]{ohto03} Ohto A., Kawano N., Fukazawa Y. 2003, PASJ, 55, 819

\bibitem[2005]{osullivan05} {{O'Sullivan}, E., {Vrtilek}, J.~M. \&  {Kempner}, J.~C.}, \ 2005 ,\apjl, 624, L77

\bibitem[2002]{pao2002} Paolillo M., Fabbiano G., Peres G., Kim D.-W. 2002,
ApJ, 565, 883

\bibitem[2005]{pointe05} {{Pointecouteau}, E. \& {Silk}, J.}, \ 2005, \mnras, 364, 654


\bibitem[1993]{1993sdgc.proc..357P} Pryor, C.~\& Meylan, G.\ 1993, in "The Structure and Dynamics of 
Globular Clusters", ASP
Conf.~Ser.~Vol.~50, eds. S.~G.~Djorgovski and G.~Meylan, p.~357

\bibitem[2004]{reed05} Reed, D., Governato, F., Verde, L. et al. 2005, MNRAS 357, 82

\bibitem[2004]{1399.2} Richtler, T., Dirsch, B., Gebhardt, K., et al.\ 2004, \aj, 127, 2094 

\bibitem[2003]{romanowsky} Romanowsky, A.~J., Douglas, N.~G.,
Arnaboldi, M. et al.\ 2003, Science, 301, 1696


\bibitem[2002]{sanders02} Sanders, R.~H.~\& McGaugh, S.~S.\ 2002,
\araa, 40, 263

\bibitem[2003]{sanders03}  Sanders, R.~H.~ \ 2003, \mnras, 342, 901

\bibitem[1974]{stephens} Stephens, M.~A.  \ 1974 Journal of the American Statistical Association, 69, 730 

\bibitem[2003]{2003ApJ...585L.121T} 
Temi, P., Mathews, W.~G., Brighenti, F., \& Bregman, J.~D.\ 2003, \apjl, 
585, L121 

\bibitem[1979]{1979AJ.....84.1511T} Tonry, J.~\& Davis, M.\ 
1979, \aj, 84, 1511 

\bibitem[2001]{2001ApJ...546..681T} Tonry, J.~L., Dressler, A.,
Blakeslee, J.~P., et al.\ 2001, \apj, 546, 681

\bibitem[1994]{marel94} {{van der Marel}, R.~P.}, \ 1994, \mnras, 270, 271

\bibitem[2004]{wegner04} {{Wegner}, G., {Bernardi}, M., {Willmer}, C.~N.~A. et al.\
2004, \aj, 126, 2268}


\bibitem[2002]{2002ApJ...579...600X} Xu, H., Kahn, S. M.,
 Peterson, J. R., et al. \ 2002, \apj, 579, 266
 
\bibitem[1996]{1996ApJ...462..266Y} Yoshii, Y., Tsujimoto, T., \&
  Nomoto, K.\ 1996, \apj, 462, 266

\bibitem[1995]{1995ApJ...447..656Y} 
Yungelson, L., Livio, M., Tutukov, A., \& Kenyon, S.~J.\ 1995, \apj, 447, 
656 

\bibitem[1997]{zhao97} {{Zhao}, H.}, \ 1997, \mnras, 287, 525

\bibitem[1939]{1939PASP...51...36Z} Zwicky, F.\ 1939, \pasp, 51, 36 
\end{thebibliography}
\end{document}